\documentclass[useAMS,usenatbib]{mn2e}
\usepackage{graphicx}

\def\la{\raise.5ex\hbox{$<$}\kern-.8em\lower 1mm\hbox{$\sim$}}
\def\ma{\raise.5ex\hbox{$>$}\kern-.8em\lower 1mm\hbox{$\sim$}}

\def\msol{M$_{\odot}$ }

\def\kms{$\rm km\, s^{-1}$}
\def\cm3{$\rm cm^{-3}$}
\def\Ts{$\rm T_{*}$~}
\def\Vs{$\rm V_{s}$~}
\def\n0{$\rm n_{0}$}
\def\B0{$\rm B_{0}$}

\def\erg{$\rm erg\, cm^{-2}\, s^{-1}$}
\def\mum{$\mu$m~}

\def\L12{L$_{12\mu m}$~}
\def\F12{F$_{12\mu m}$~}

\def\Hb{H${\beta}$~}
\def\Ha{H${\alpha}$~}
\def\Haa{H${\alpha}_{calc}$~}
\def\Ly{Ly$\alpha$~}
\def\La{L$_{H\alpha}$~}

\title[Galaxies at redshift 0.001$<$z$<$3.4]{Physical conditions and element abundances 
in  galaxies   at redshift  0.001$<$z$<$3.4 : evolution trends}
\author[M. Contini]{M. Contini  \\
School of Physics and Astronomy, Tel Aviv University, Tel Aviv 69978, Israel \\
}

\begin{document}

\date{Accepted: Received ; in original form 2010 month day}

\pagerange{\pageref{firstpage}--\pageref{lastpage}} \pubyear{2009}

\maketitle

\label{firstpage}

\begin{abstract}
We have collected an heterogeneous sample of galaxies at redshifts  0.001$\leq$z$\leq$3.35 
with the aim of  exploring the evolution  of the physical parameters and of the
element  abundances calculated by  modelling  the observed spectra.
The evolution picture  which results from our calculations  shows that:
1) the emitting gaseous clouds in both active galactic nuclei (AGN) and starburst (SB) galaxies 
have  a density broad peak at z$\sim$0.5 becaming
 less dense and more elongated  with time.
2) The \Ha flux emitted from the AGN clouds has a maximum at z$\sim$ 0.1.
3) The observed star temperatures in SBs increase with time, indicating younger stars in local galaxies.
4) The  O/H relative abundances  show splitting between solar and $\geq$0.1 solar 
at 0.3$<$z$<3$. N/H also shows splitting in this z range but it also has a broad flat
maximum between z=0.1 and z=0.001. 
5) The AGN narrow line region  and the regions  surrounding  the SB show fragmentation
at  z $\geq$0.35.
6) Metallicity (in terms of O/H)
correlates with  \Ha luminosity \La (and therefore with the star formation rate)  
with different trends  regarding SBs and
AGNs for \La $\geq$ 10$^{41}$ erg s$^{-1}$. A rough correlation is seen also for N/H. 
There seems to be no correlation for low luminosity galaxies.

\end{abstract}


\section{Introduction}

The evolution with  redshift of   galaxy  physical characteristics  and of
the   star formation rates (SFR) 
has been recently investigated  in   different  types of galaxies (Teplitz et al. 2003, Caputi et al. 2007,
Stierwalt et al 2013, Straughn et al. 2009, Shim et al. 2009, etc. ).

The SFR is calculated from the  \Ha luminosity (\La) or from the infrared luminosity (L$_{IR}$), depending on the data 
available from the observations.
\La therefore, together with the  line intensities and profiles, gives us the information about the physical
structure of the interstellar medium which is important to the formation of stars
 (Spaans \& Carollo 1997).
 The  mutual feedback of starbursts (SB) 
and  active galactic nuclei (AGN)   leads to an enhanced \Ha luminosity  from the galaxy
 because both the SB and the   AGN  contribute.
The  \Ha emission line flux  and the  flux  of  lines corresponding to  the other elements 
emitted  from  the gas  close to the SB and  
 from  the gas  ionized by the  AGN radiation flux  are summed up 
in the  characteristic  spectra observed at Earth (Contini 2013a).

Regarding the epoch of star forming activity,
most  stars in  massive galaxies (M$>$10$^{11}$ \msol) formed $\sim$  8-10 billion 
years ago,  when star formation was at its  maximum (Swinbank  et al 2012).
It is believed that the optimal metallicity (the metal production rate is a direct measure 
of the star formation rate) corresponds to a period of intense star formation
at about 1$<$z$<$2 (Spaans \& Carollo 1997, Madau et al. 1996).

The  abundances of the heavy elements in the surrounding of the starbursts
are affected not only by the star activity at different ages, but also  by  galaxy interactions.
The  discovery of  AGNs  showing a double nucleus (e.g. Fried \& Schultz 1983
for NGC 6240, Fabbiano et al 2011 for NGC 3393, etc) demonstrated that some local objects are the
result of merging.
Collision of galaxies at higher redshifts  yields  star formation
by compressing, collisionally heating and ionizing  the gaseous clouds throughout the galaxy.
For instance, based on FIR data from the Herschel Space Observatory, Santini et al (2012)
gave evidence of  an enhanced star formation activity in the host galaxies of X-ray -selected AGN
at 0.5$<$z$<$2.5 as compared to non-active galaxies. 
X-rays may reveal high temperature  gas downstream of high velocity shocks.

It is suggested that mergers and tidal interactions
lead to a low O/H in the galaxy central region and that
a high  N/O  could reveal   a powerful starburst (Pustilnik et al. 2004).
In a sample of star-forming luminous compact galaxies showing recent interactions at z $<$ 0.63,
Izotov et al. (2011)
found  significant  scattering in N/O values, with  probably enhanced nitrogen abundance (see also
Contini et al 2012 regarding the local galaxy NGC 7212).

Metallicity is represented  by the O/H  relative abundance because oxygen is the most abundant of the   heavy elements
and  the  corresponding spectral lines are the strongest e.g. in the optical range.
The N/O abundance ratio as a function of  O/H  is considered to be  useful
in determining the chemical evolution of galaxies (Edmunds \& Pagel 1978; Torres-Peimbert et al. 1989;
Vila Costas \& Edmunds 1993; Izotov et al. 2006).
Investigating the correlation between relative abundances and galaxy morphologies of blue compact galaxies
at 0.20$\leq$z$\leq$0.35, Chung et al (2013)   found that
 the oxygen and nitrogen abundances
may be   related to the local environments of the galaxies.

Metallicity (in terms of O/H)   is calculated from the
[OIII]/\Hb and [OII]/\Hb line ratios.
In an accompanying paper (Contini 2013b) we discuss
the reliability of the results obtained by modelling spectra showing only a few oxygen to \Hb  line ratios
 on the basis of the Kakazu et al (2007)  ultrastrong emission line
of extremely low metallicity  galaxy sample at z$\sim$ 1 and the Xia et al (2012)  faint galaxy
 sample at 0.6$\leq$z$\leq$2.4.

At low redshifts (z$\leq$0.1)  the calculated  relative abundances   are constrained by
   many lines from different elements in different ionization levels  throughout  UV-optical-infrared spectra, while
the  modelling  of spectra  at higher z
is prevented when one or more key lines from both  high and low ionization levels are leaking (see Contini 2013b).
Therefore, at higher  redshifts (z$>$0.1)
 direct  methods are generally adopted  to determine O/H.
The "direct" method  (Kakazu et al. 2007,  Isotov et al 2011),  based on Seaton (1975) and  Pagel et al (1992) works,
  requires  three oxygen lines ([OIII]4959,5007, [OII]3727).

We use composite models  accounting for the combined effect of photoionization from the AGN or from the SB and shocks
(e.g. Contini 2013a and references therein) which are  adapted to   deal with the collisional processes throughout
the   mergers.
Therefore, the choice of the models will be constrained by  the FWHMs of the line profiles which give
a hint about the velocity field.

 The mutual interaction between the nuclear activity and the starbursts  in the same object
is  investigated in this paper by  two different kinds of models :
those  characterised by a power-law  photoionising  flux
 and those  characterised by a black-body radiation flux.
 Both  photoionization and  shocks are accounted for in the calculation
of the spectra.

The analysis of  the spectra observed  at different  locations throughout   the
narrow emission line region (NLR) and beyond it in single AGN, allows to 
 calculate   the relative abundances of the heavy elements to H and the ages  of
processes, such as star formation  in close or far regions  from the AGN (Ishibashi et al 2013). 
The  location of the collision zone   is  revealed by the shock velocity distribution throughout the galaxy
(Contini et al 2012, Contini 2012a,b).

We have collected some samples of galaxy spectra at different redshifts   0.001 $<$z$<$ 3.4.
The  composed sample  accounts 
for active and non-active galaxy samples and individual objects, 
including  for instance  X-ray selected AGN,  ultraluminous infrared galaxies (ULIRGs),  
damped \Ly absorber  (DLA) galaxies,  Seyfert galaxies,  3CR radio galaxies, 
star-forming galaxies,  low ionization nuclear emission regions (LINERs),  SB galaxies, etc.,
   observed with different angular and spectral resolutions. 
The galaxies were selected on the basis of their spectra, i.e. showing enough lines to yield a reliable modelling.
Therefore, the  composed sample  is  relatively  poor in number of objects.
Unfortunately, far away objects characterised by fragmented gaseous clouds  with  
 low  densities  and  metallicities  have been most probably excluded.
Moreover, the
 observed  line ratios which provide the tools for calculating the
physical and chemical conditions in galaxies  at each redshift, 
 are often uncertain due to selection effects, calibration biases and  different 
corrections from obscuration (Gunawardhana et al. 2013).

 Samples of galaxies belonging to a certain class (e.g. active galaxies) and  observed by similar techniques, 
generally cover
restricted  z ranges  and  show  nearly the same  lines  for each object spectrum. In an accompanying paper
(Contini 2013c, hereafter Paper I) we have investigated the sample of active galaxies observed by Ramos Almeida et al. (2013).
The results of modelling show that  the  parameters exhibit some  smooth 
or varying trends in the redshift range 0.4$<$z$<$1.15. 
However, their interpretation is limited to a restricted although significant redshift range.

We believe that the results obtained by  modelling  the  line   spectra emitted from the  gas  
within  different types of galaxies will show the trends of  the  physical parameters 
such as e.g. density,  shock velocity, star temperature, etc. 
and of the element abundances  throughout the  z range, independently from the object identity.
Moreover, adopting an heterogeneous  sample of galaxies, smooth profiles or  well defined  fluctuations    
throughout  a relatively extended redshift range can be  better  recognised
 and interpreted in the light of galaxy  evolution and age.

The samples gathered in this work are the following.
At  redshift ranges  1.1 $\leq$ z $\leq$ 3.35 we have analysed  the spectra observed by Brand et al. (2007) from
optically faint ultraluminous IR galaxies (ULIRGs).
The results of  modelling   Ramos Almeida et al. (2013) galaxy spectra covering the redshift range 0.4$<$ z $<$ 1.15,
which were presented  in Paper I, are accounted for.
Recently, the spectrum of the damped \Ly absorber galaxy  (DLA)
 in the z=2.93 QSO 2222-0964 has been presented by Krogager et al (2013).
It has been added to our analysis.
We   included  also  the line spectrum observed by Schirmer et al (2013)  from
the galaxy J4022-0927 at z=0.326, because it  covers  the near UV-- near IR range.

Our aim is also  to find out and  compare the physical conditions  of  galaxies at   lower redshifts.
The star-forming region spectra, e.g.
 the relic radio-galaxy at z$<$0.165 presented by Capetti et al (2013)
and the  spectra observed in the HII regions of  compact star forming galaxies at 0.11 $<$ z $<$ 0.5 by
Kobulnicky \& Zarisky (1999, hereafter KZ99) are analysed.

To  learn  about the evolution of the characteristic parameters
of active galaxies (both AGN and SB) on a large scale, we  have extended the  spectral sample  to local galaxies,
namely,  AGNs and SBs, such as those included in the sample of the low ionization nuclear emission
line regions (LINERs) by Contini (1997) and in
the  sample of starburst galaxies by Viegas et al (1999), as well as the local merger Seyfert galaxies
(Contini 2012a,b, 2013a).

Our sample is not complete, so, in order to cover the redshift range by a continuous distribution of the 
different parameters, we have
included the data obtained from  a subsample of  galaxy spectra presented by Winter et al. (2010).
We have selected  the most  observed  objects and we have modelled the line ratios.

In Fig. 1 the  line ratios observed from the selected galaxies are  plotted  throughout a Baldwin-Phillips-Terlevich (BPT)
diagram (Baldwin et al. 1981). 
Fig. 1 shows that except for the HII region  sample by KZ99 all the ensembles of  line ratios within the other samples 
cross at least one of
the lines separating AGNs from HII regions. The grids of calculated line ratios for local merger galaxies
 by Contini (2012a, fig. 1) show that the  domains of AGN and SB in a BPT diagram may overlap.

\begin{figure}
\includegraphics[width=8.6cm]{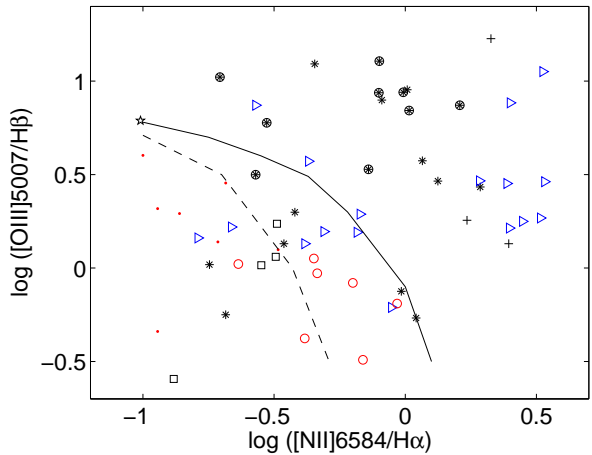}
\caption{The distribution of the  galaxies throughout a BPT diagram.
 Ramos Almeida et al. (2013): black asterisks; Brand et al (1997): black squares; Krogager et al (2013): black pentagram;
KZ99: red dots; Winter et al (2010): black filled circles; Schirmer et al (2012): black asterisk;
local mergers (Diaz et al 1988); Fosbury \& Wall  (1979); Veilleux et al. (1999) : black +;
Capetti et al (2013): red circle; Contini (1997): blue triangles; Contini \& Contini (2003) : red circles;
the black dashed and solid lines represent the empirical separation between AGN and HII regions by Kewley et al (2001)
and Kauffmann et al (2003), respectively
}
\end{figure} 

The results of modelling the composed sample are presented in  Sect. 2  and discussed in Sect. 3 
in the light of evolution with z.  Concluding remarks follow in Sect. 4.

\section{Modelling results}

To calculate the spectra we
 have adopted the code  {\sc suma}\footnote{http://wise-obs.tau.ac.il/$\sim$marcel/suma/index.htm}, which
 simulates the physical conditions in an emitting gaseous cloud under the coupled effect of
photoionization from an external radiation source and shocks 
(see Paper I  for a detailed description of the code).

The input parameters which characterise the model are : the  shock velocity \Vs, the atomic preshock density \n0,
the preshock magnetic field \B0. They define the hydrodynamical field and  are  used in the calculations
of the Rankine-Hugoniot equations  at the shock front and downstream. These equations  are combined into the
compression equation which is resolved  throughout each slab of the gas, in order to obtain the density
profile downstream and consequently, the temperature of the gas.
We adopt  for all the models \B0=10$^{-4}$ gauss. 
The input parameter  that represents the radiation field  in AGNs is the power-law
flux  from the active centre $F$  in number of photons cm$^{-2}$ s$^{-1}$ eV$^{-1}$ at the Lyman limit.
The spectral indices are $\alpha_{UV}$=-1.5
and $\alpha_X$=-0.7.
If the radiation flux  is  a black body radiation from the stars,
the input parameters are the colour  temperature of the  star \Ts
and  the ionization parameter $U$
(in  number of photons per  number  of electrons at the nebula).

The geometrical thickness of the emitting nebula $D$,
the dust-to-gas ratio $d/g$, and the  abundances of He, C, N, O, Ne, Mg, Si, S, A, and Fe relative to H
are also accounted for.

In our models the flux from an external source can reach the  shock front (the inflow case,
indicated by the parameter $str$=0) or the edge of the
cloud opposite to the shock front  when the cloud propagates outwards from the AC or from the starburst
(outflow, indicated by $str$=1).

The observed line ratios presented in the following  are all reddening corrected. 

\begin{table*}
\caption{Comparison of calculated with observed galaxies from the Brand et al (1997) sample at 1.37$\leq$z$\leq$3.3}
\small{
\begin{tabular}{lccccccccccccccc} \hline  \hline
\       &  z   & [OIII]/\Hb & [NII]/\Hb     & FWHM  & \Hb  & \Vs & \n0  &log$F$ &   $U$ & \Ts   & N/H     & O/H       & $D$           & $str$\\
\       &         &            &          & \kms & \erg & \kms& \cm3 & - &  -  & 10$^4$ K & 10$^{-5}$&10$^{-4}$  & 10$^{15}$cm  &  \\ \hline
\  J$^1$ & 1.37   & 0.34       & 0.38      & 375  &  $<$17.6 10$^{-18}$&-& -  &   -      &  -  & -  &  -  & -   & -  &- \\
\ m$_{AGN}$&  -    & 0.46      & 0.28     &   -  & 0.028   & 390& 500 & 10     &  -    & -  &  3.& 2.6& 2.2&0 \\
\ m$_{AGN}$& -     & 0.5      & 0.5      &   -  & 0.02    & 390& 500 & 10     &  -    & -   & 3. & 2.6& 2.5&1 \\
\ m$_{SB}$& -     & 0.32       & 0.3      & -    &0.018    & 390 & 300& -      & 0.01   &3.  & 7. & 3. & 10&1  \\
\  J$^2$ &2.11   & 2.3         & 0.94       & 469  &$<$ 11.9 10$^{-18}$&  -  & - &-       &-      &-   &-      &-      &-   &-   \\
\ m$_{AGN}$ & -   & 2.         & 0.93      &  -   & 0.013      &450  & 300& 10     & -   & -  & 3.5& 6.6& 1.3&0  \\
\ m$_{AGN}$ & -   & 2.23       & 1.        & -    & 0.01      & 450  & 210& 10     & -   & -  & 5. & 6.6& 1.3&1 \\
\ m$_{SB}$ & -   & 2.3         & 0.86       & -    & 0.0038    & 480  & 320 & -     & 0.002& 4.& 8.& 6. & 2.3&1 \\
\  J$^3$   & 2.18& 1.53        & 0.93       & 1617 & $<$42.9 10$^{-18}$&  -  &  -  &  -   &  -    &  -  &  -   &  -    &   -  &- \\
\ m$_{AGN}$ &  -  & 1.48       & 0.86      & -    &0.028       &1600 & 210 &  10   & -   &  -  & 9.&6.6 & 1.&0  \\
\ m$_{AGN}$ &  -  & 1.44       & 0.8       & -    & 5.7        &1600 & 330 & 10    & -   &  -  &11. &7.6& 1000.&1  \\
\ m$_{SB}$ &  -  & 1.43        & 0.8        & -    & 5.7        &1600 & 330 &  -    & 0.01& 3. & 11.&7.6&1000.&1  \\
\ J$^4$    & 2.223&1.38        & 0.82       & 585  & $<$59.8 10$^{-18}$& -   &  -  &  -   &  -    &  -   &  -    &   -  &  -  &-  \\
\ m$_{AGN}$ & -    & 1.36      & 0.85      & -    & 0.014      &680  & 250 & 9.78  & -   &  -   & 7. & 6.&3.3&0 \\
\ m$_{AGN}$ & -    & 1.4       & 0.82      & -    & 0.001      &650  & 250 & 9.6   & -   &  -   & 5. & 6.&3.3&1 \\
\ m$_{SB}$ & -    & 1.45       & 0.72       & -    & 0.007      &600  & 250 & -     & 0.004& 2. & 10. & 5.&12.&1 \\
\ J$^5$    & 3.355& 1.87 $^6$  &  -         & 533  & $<$57.9 10$^{-18}$& -   &  -  & -     & -    &  -   &  -    &  -   &  -  &-  \\
\ m$_{AGN}$ & -    & 1.65      & 0.6       &  -   & 0.009      &550  & 250 & 9.6   &  -   &  -   & 10. & 6.& 3.3&0 \\
\ m$_{AGN}$ & -    & 1.44      & 0.85      &  -   &0.0083      &550  & 250 & 9.6   &  -    &  -   & 10. & 6.& 5.3&1  \\
\ m$_{SB}$ & -    & 1.45      & 0.72       &  -   &0.007       &600  & 250 &  -    & 0.004 & 2. & 10. & 5.& 12.&1 \\ \hline

\end{tabular}}

$^1$ J143027.1+344007 ; $^2$ J143011.3+343439 ; $^3$ J142842.9+342409 ; $^4$ J142800.7+350455 ; $^5$ J142644.3+333051 ;

$^6$ not  extinction corrected;

$F$ is in  photons cm$^{-2}$ s$^{-1}$ eV$^{-1}$ at the Lyman limit ;

\end{table*}

\begin{table*}
\small{
\caption{Comparison of calculated with observed line ratios to \Hb for the DLA galaxy (Krogager et al 2013) at z=2.35}
\begin{tabular}{lccccccccccccccccccc} \hline  \hline
\          & \Ly   &   [OII] & \Hb&[OIII]& \Ha  & [NII]  &  \Hb &\Vs   & \n0  & log$F$   & $U$ & \Ts    & N/H  & O/H   & $D$ & $str$   \\
\          &  -    &   -     &  - &  -   &  -   &   -    & \erg & \kms & \cm3 &- & -   & 10$^4$ K  &10$^{-5}$&10$^{-4}$& 10$^{18}$cm&-\\ \hline  \hline
\  obs     & 10.1  &   1.6   &   1& 8.2  & 2.86 &  0.284 &-     &  -   &  -   &  -       & -   & -      &  -    &  -   &  - &-    \\
\ M$_{AGN}$ & 26.9   &  2.8   &   1& 8.   & 2.9  & 0.4    &0.21  & 130  & 400  & 10.8    & -   &  -     & 2. &5. & 1.&0     \\
\ M$_{SB}$ & 26.7  &   1.8   &   1& 8.24 & 2.9  & 0.48   &0.075 & 130  & 450  & -        & 0.09&4.5   & 7. &7.9& 0.016&1    \\ \hline        
\end{tabular}}

$F$ is in  photons cm$^{-2}$ s$^{-1}$ eV$^{-1}$ at the Lyman limit

\end{table*}

\subsection{Spectra from the  Brand et al (2007) galaxy sample  at 1.37 $\leq$ z $\leq$ 3.35}

In Table 1 we present the modelling of the spectra  of optically faint ULIRGs
 at relatively high z reported by Brand et al. (2007). Their sample of 10 galaxies was selected from a 24 \mum survey of the
NOAO Deep Wide-Field Survey (NDWFS; Jannuzi \& Dey 1999) Bo\"{o}tes field. The sample was observed by the NIRSPEC (McLean et al 1998)
on the Keck II telescope or the NIRI (Hodapp et al 2003) on the Gemini North telescope.

We have reproduced the  [NII]6583/\Hb and [OIII]5007/\Hb
line ratios by models referring to AGNs
(m$_{AGN}$) as well as by models referring to the starburst (m$_{SB}$).
In the former case we have used both models showing clouds
inflowing towards the AGN ($str$=0)  and outflowing ($str$=1)  from the galaxy. For starbursts, we consider that only the outflowing case
makes sense.
Most of the line observed fluxes are upper or lower limits. Moreover, the observed lines are not enough to distinguish
between  SB dominated or  AGN dominated models. On the other  hand, with the help of the FWHM of the forbidden lines,
we could constrain the dynamical and therefore also the radiation parameters.
We can thus  obtain  the distribution of the
physical and chemical parameters  at relatively high redshifts. 

The observed line ratios in Table 1 are  corrected for extinction
adopting E(B-V) values which appear in Brand et al (2007, table 1).
The modelling results show that the photoionizing flux from the active centre does not exceed the values found for
low ionization AGNs even at relatively high z. Perhaps this happens because Brand et al galaxies are ULIRGs
and the high dust-to-gas ratio characteristic of ULIRGs (Contini \& Contini 2002) 
can  reduce the  intensity  of the flux reaching the clouds.
Moreover the cloud geometrical thickness is low indicating strong fragmentation most probably due to turbulence
related to  the relatively high  velocity shocks. In the case of SB, the temperature of the stars is rather low,
as well as the ionization parameter.

\subsection{The spectrum of the DLA galaxy at z=2.35 towards QSO, SDSSJ2222-0946}

The emission line spectrum from  the  DLA galaxy at z=2.35 observed by Krogager et al. (2013) by  VLT/X-shooter, shows the
\Ly, [OII]3727+, [OIII]5007+,
[NII]6584 besides \Ha and \Hb lines. The results of modelling the extinction corrected line ratios  using both an AGN model and a SB model
are shown in Table 2. The  \Ly/\Hb line ratio is overestimated by a factor $\geq$ 2 by the models, however, lower values 
would lead to very low [OII]/\Hb, [OIII]/\Hb  and [NII]/\Hb. This indicates that many different conditions contribute to the observed spectrum.
   We derived O/H = 5 10$^{-4}$ and 7.9 10$^{-4}$, 
(while Krogager et al. 2013 found O/H=2.5 10$^{-4}$)
and N/H
= 2 10$^{-5}$ and 7 10$^{-5}$ for AGN and SB models, respectively, 

 The shock velocities deduced from the FWHM of the line profiles
are rather low (\Vs $\sim$ 130 \kms) and the preshock densities (\n0 $\sim$ 400 \cm3) are
 consistent with those   found for other  Ramos Almeida et al objects in Paper I.

\subsection{The line spectrum of the  Seyfert galaxy J2240-0927 at z=0.326}

We would like to   include  to the   analysis of the physical conditions in  galaxies at z$\leq$0.5
  the spectrum of one of the "Green Been" prototype galaxies observed by Schirmer et al (2012),
 J2240-0927 at z=0.326, which shows
 one of the most luminous NLR of Seyfert 2 galaxies. 
 Characteristic of this galaxy is the extended NLR covering more than  26 $\times$ 44 kpc$^2$ and
the large region surrounding   the NLR.
The spectrum is based on observations made with the ESO Telescopes at La Silla and Paranal
Observatories in Chile,  and observations obtained with MegaPrime/MegaCam and by the Gemini Observatory.

The  ratios of selected emission lines (corrected for extinction) to \Hb are reported in Table 3.
Schirmer et al suggest velocities between 200 and 300 \kms calculated from the FWHM of the line profiles.
We will constrain the models adopting  this velocity range.
The spectrum is rich of lines from the near UV - optical-  near IR ranges.
The [NeIII]/\Hb line ratio is overpredicted by the models  accounting for the summed $\lambda$3689 and $\lambda$3956 line components.
However, a high [NeV]/\Hb cannot be obtained by models MJ$_{200}$ and MJ$_{600}$
which show a satisfactory agreement for all the other lines.
A lower flux must be adopted. In the case of a model corresponding to \Vs=600 \kms,
a lower $F$ reduces too much the [OIII]5007+/[OI]6300+ line ratio. A better fit to the
[NeV]/[NeIII] results adopting both a low flux and a low \Vs.
The models which refer to the outward  motion of the NLR clouds   better explain 
the observed line ratios than those referring to inflow.

We have found that N/H (7. 10$^{-5}$) is slightly lower than solar (10$^{-4}$), O/H and Ne/H are solar and Mg/H=
 6.10$^{-6}$ while the solar value is 2.6 10$^{-5}$.

\begin{table}
\caption{Comparison of calculated with observed line ratios for the galaxy J2240-0927 at z=0.326}
\begin{tabular}{lcccccc} \hline  \hline
\  line        & obs    & MJ$_{200}$&MJ$_{600}$&MJ$_{lF}$&MJ$_{lFlV}$\\ \hline
\ MgII 2789    & 0.24     & 0.3   & 0.34&0.4 &0.18\\
\ [NeV] 3427   & 0.22     & 0.06  & 0.06&0.2 &0.13 \\
\ [OII] 3727+  & 3.72     & 4.4   & 2.6  &3.& 3.9\\
\ [NeIII] 3869+& 0.71     & 1.0   & 1.2  &0.96& 1.2\\
\ H$\gamma$    & 0.40     & 0.46  & 0.46 & 0.46 &0.46\\
\ [OIII] 4363  & 0.13     & 0.07  &0.08  &0.05 & 0.09 \\
\ HeII 4686    & 0.13     & 0.27  &0.27  & 0.15 & 0.27\\
\ \Hb          & 1        & 1    & 1     & 1  & 1 \\
\ [OIII] 5007+ & 10.56    & 9.8   & 10.9 & 1.84& 9.83\\
\ HeI 5876     & 0.09     & 0.12  & 0.12 & 0.18&0.11\\
\ [OI] 6300+   & 0.55     & 0.3   & 0.4  &1.65 &0.13$^1$\\
\ \Ha          & 4.25     & 2.9   & 2.9  &3. & 2.9\\
\ [NII] 6584+  & 2.36     & 2.9   & 2.7  & 3.9&2.9\\
\ [SII] 7617+  & 2.11     & 1.5   &0.7   & 1.3&0.75\\
\ [ArIII] 7138+& 0.11     & 0.14  &0.15  & 0.1&0.16\\
\ [SIII]9071+  & 1.71     & 2.07  &2.4   &2.1 &1.5\\
\ HeI 10833    & 0.30     & 0.65  & 0.96 & 1.09 &0.6\\ \hline
\ \Hb abs$^2$     &  -      &  0.057&0.173 & 0.05&0.03\\    
\  Vs (\kms)    &  -      & 200.  &600. & 600. &300.\\
\ \n0 (\cm3)    &  -      & 390.  &360. & 360.&280.\\
\ $F$ (10$^{10}$ units$^3$)& -      & 3. &9.& 2.&2.\\
\ $D$ (10$^{15}$ cm)       &  -     &6. &6.& 6.&6.\\
\ n (10$^3$ \cm3)       &  -     &10     &28.& 28.&9.3\\
\ N/H (10$^{-5}$)           &  -     &7.   &7.&7.&7.\\
\ O/H (10$^{-4}$)           &  -     &6.6 &6.6&6.6& 6.6\\
\ Ne/H (10$^{-4}$)          &  -     & 1. &1. & 1.&1.2\\
\ Mg/H (10$^{-6}$)          &  -     & 5. &5. & 5.&5.\\
\ S/H  (10$^{-5}$)          &  -     & 1.6&1.6& 1.6&2.\\
\ Ar/H (10$^{-6}$)          &  -     & 1.3&1.3& 1.3&1.3\\
\ str            &  -     & 1     &1     &1      &1     \\ \hline

\end{tabular}

$^1$ +[SIII]6312; $^2$ \Hb calculated absolute flux in \erg; 
$^3$ in photons cm$^{-2}$ s$^{-1}$ eV$^{-1}$ at the Lyman limit

\end{table}

\begin{table}
\caption{Comparison of calculated with observed line ratios for 3C 258 at z=0.165}
\begin{tabular}{lccccccc} \hline  \hline
\             & obs   & model \\ \hline
\ \Hb         & 1.   &   1.  \\
\ [OIII] 5007+ & 1.4  & 1.45 \\
\ [OI] 6300+   & 0.27 & 0.2 \\
\ [NII] 6548+  & 0.67 & 0.6 \\
\ [SII] 6717   & 0.87 & 0.73 \\
\ [SII] 6731   & 0.53 & 0.7  \\
\ \Vs  (\kms)  & -    & 300  \\
\ \n0  (\cm3)  & -    & 40   \\
\ \Ts (10$^4$ K)      & -    & 4.7\\
\ $U$          & -    & 0.02\\
\ $D$ (10$^{19}$ cm)     & -    & 1.\\
\ N/H (10$^{-5}$)         & -    & 3.\\
\ O/H (10$^{-4}$)         &-     & 7.\\
\ S/H (10$^{-5}$)         & -    & 1.\\
\ \Hb (\erg)   &  -   & 0.0054\\
\ n  (\cm3)   &  -   & 506.   \\ \hline

\end{tabular}
\end{table}

\subsection{Relic radiogalaxies (z$\leq$0.3)}

Capetti et al (2013) presented the optical spectra of 3CR radiogalaxies at z$<$ 0.3.
which they interpreted as relic AGNs. The only object of the sample which  shows data suitable
to  modelling is 3C 258. The line ratios are characteristic of star-forming regions, therefore
we have modelled the spectrum by a starburst model. The results appear in Table 4.
Capetti et al claim that the object is very compact for its redshift and the arc-like structure at SE
suggests a recent merger. O/H is about solar and N/H is 0.3 solar.

\begin{table*}
\caption{Comparison of calculated with observed line ratios to \Hb for the KZ99 sample}
\begin{tabular}{lccccccccccccccc} \hline  \hline
\ galaxy          &[OII] & [NeIII] & H$\gamma$& [OIII] & HeI    & HeII  & [OIII] & HeI   &[SIII] & \Ha &[NII] & [SII] & [SII]& z\\
\                 & 3727+& 3869+     & 6340   & 4363   & 4471   & 4686  & 5007+  & 5876  &6312+  & 6564& 6584+ & 6717  &6731 & \\ \hline
\ {\tiny L2-408115}&3.8  &$<$0.1   &0.43      &$<$0.025&$<$0.03 &$<$0.024&1.67   &0.12   & $<$0.020     & 2.92& 0.95  &0.6    &0.45 &0.197\\
\ cal1            &4.2   &0.13     &0.46      &0.03    &0.05    0.0002   &1.55   &0.14   & 0.055   &2.96 &0.978  &0.637  &0.47 & \\
\ {\tiny L2-410083}&0.6  &0.51     &0.48      &0.15    &0.038   &$<$0.008&8.88   &0.106  &$<$0.005      &2.86 &0.024  &0.064  &0.051&0.109 \\
\ cal2            &0.60  &0.46     &0.46      &0.04    &0.046   &0.38    &8.1    &0.1    &0.007         &2.9  &0.028  &0.04   &0.04 &  \\
\ {\tiny L2-411500}&3.38  &0.36     &0.46      &0.031   &$<$0.1  &0.012   &4.485  &0.09  &$<$0.06       & -   & -     & -     & -   &0.193\\
\ cal3            &4.1   &0.39     &0.46      &0.06    &0.09    &0.00063  &4.6    &0.13   &0.068         &2.9  &1.1    &0.25   & 0.23&     \\
\ {\tiny SA 68}$^a$       &1.93 &0.127    &0.44      &$<$0.012&$<$0.011&$<$0.011&2.77   &0.07    &$<$0.023      &2.87 &$<$0.33&0.406  &0.386&0.285\\
\ cal4            & 2.5  &0.2      &0.46      &0.03    &0.048   &0.0003  &2.84   &0.13   &0.1           &3.   & 0.38  &0.37   &0.38 &    \\
\ {\tiny SDG 125}         &3.67  &0.27     &0.46      &$<$0.014&0.036   &$<$0.01 &3.14   &0.13   &$<$0.003      &-    & -     &  -    & -   & 0.3989\\
\ cal5            &3.8   &0.21     &0.46      &0.046   &0.047   &0.0003   &3.66   &0.136  &0.035         &3.   & 0.46  &0.14   & 0.15& \\
\ {\tiny SDG 146}         &1.94  &$<$0.07  &$<$0.36   &$<$0.051&$<$0.26 &$<$0.025&1.48   &$<$0.1 &$<$0.28       &-    & -     & -     & -   & 0.4006 \\
\ cal6            & 2.2  &0.12     &0.46      &0.019   &0.047   &0.00028  &1.58   & 0.14  & 0.02         & 3.  &0.38   &0.11   & 0.13&   \\
\ {\tiny SDG 173A}        & 3.4  &$<$0.1   & 0.484    & $<$0.065& $<$ 0.1&$<$0.09&3.8    &$<$0.17 &$<$0.4  & $<$0.39 &$<$0.6&$<$0.16&$<$0.16 &0.3996\\
\ cal7            &3.8   & 0.21    & 0.46     & 0.046   & 0.047  &0.0003  &3.66  & 0.136    &0.035    & 3.     &0.46     &0.14  &0.15 & \\ \hline

\end{tabular}

$^a$ SA 68-206134

\end{table*}

\begin{table*}
\caption{Comparison of calculated with observed line ratios to \Hb for the KZ99 sample}
\begin{tabular}{lccccccccccccccc} \hline  \hline
\ object          &[OII] & [NeIII] & H$\gamma$& [OIII] & HeI    & HeII  & [OIII] & HeI   &[SIII] & \Ha &[NII] & [SII] & [SII]& z\\
\                 & 3727+& 3869+     & 6340   & 4363   & 4471   & 4686  & 5007+  & 5876  &6312+  & 6564& 6584+ & 6717  &6731 & \\ \hline
\ {\tiny SDG 183} &3.    &$<$0.06  &0.411     & $<$0.033&$<$0.037&$<$0.036&2.61  &$<$0.043&$<$0.46&2.48&$<$0.4 & $<$0.03&$<$0.03 & 0.3994\\
\ cal8            &3.5   &0.09     &0.46      & 0.03    &0.04    &5.e-4 &2.74    &0.13    &0.01   & 3.  &0.44  & 0.04   &0.04   & \\
\ {\tiny SDG 223} &1.78  & 0.243   &0.46      &0.027    &0.035   &$<$0.01&5.34   &0.12   &$<$0.04&2.9  &0.29  & $<$0.09&$<$0.09 & 0.3996\\
\ cal9            &1.4   & 0.25    &0.46      &0.012    &0.05    &0.02   &5.3    &0.15   &6.e-5  &2.95 &0.34  &0.08    &0.08    &    \\
\ {\tiny SDG 231} &3.53  &$<$0.15  &0.44      &$<$0.015 &$<$0.014&$<$0.045&0.61  &$<$0.06&$<$0.189&2.86&$<$0.33&$<$0.16&$<$0.16 & 0.3989\\
\ cal10           &3.2   & 0.11    & 0.45     & 0.04    & 0.08   & 0.002  &0.63  & 0.26  &4.e-4   &3.1 & 0.28  & 0.16  & 0.16   &  \\
\ {\tiny L2-408570}&3.4  &$<$0.144 & $<$0.304 &$<$0.12  & $<$0.09&$<$0.09 & 2.05 &   -   &  -     & -  &  -    &  -    &  -     & 0.496\\
\ cal11            &4.   &0.16     & 0.43     & 0.15    & 0.15   & 0.004  & 1.9  & 0.5  & 3.e-3   & 3.56& 0.1  & 0.05  & 0.04   &   \\
\ {\tiny L2-408939}&3.03 &$<$0.009 &0.47      & 0.078   &$<$0.07 & $<$0.06& 3.55 &$<$0.27&$<$0.058&-    &-     &-      &-       & 0.425 \\
\ cal12            &2.3  & 0.06    & 0.46     & 0.02    & 0.05   & 0.002  & 3.7  &0.15   &0.002   & 2.94&0.12  & 0.03  & 0.024  &   \\
\ {\tiny SA 68}$^a$&2.   & $<$0.05 & $<$0.316 & $<$0.095&$<$0.07 & $<$0.06& 0.85 &$<$0.93&$<$0.208& -   &-     &-      &-       & 0.4311\\
\ cal13            & 2.2 &0.046    & 0.46     & 0.015   & 0.06   & 0.002  & 0.91 & 0.19   &4.e-4  & 2.99& 0.28 & 0.2   & 0.2    &    \\
\ {\tiny SA 68}$^b$&2.54 & $<$0.09 & 0.727    & $<$ 0.084& $<$0.08& $<$ 0.077&1.84&$<$0.12&$<$0.133& 2.09& $<$0.56 & 0.83& 0.755& 0.2343 \\
\ cal14            & 2.4 & 0.05    & 0.46     & 0.013    & 0.06   & 0.0013   &1.81&0.17&0.001&2.98 & 0.48& 0.67   & 0.73  \\ \hline 

\end{tabular}

$^a$ SA 68-206895; $^b$ SA 68-207213

\end{table*}

\begin{table*}
\caption{The physical parameters and relative abundances for  the models presented in Tables 5  and 6  for the KZ99 sample}
\begin{tabular}{lcccccccccccccc} \hline  \hline
\ model &  \Vs   &  \n0 &  \Ts      &$U$   & $D$   & N/H        &O/H       & Ne/H     & S/H      & \Hb abs & n$_H$   & z \\
\       &  \kms & \cm3 & 10$^4$ K    & -     &10$^{17}$ cm    &10$^{-4}$  &10$^{-4}$ &10$^{-4}$ & 10$^{-5}$ & \erg    & \cm3    & - \\  \hline 
\ cal1  & 50     &60    &4.          & 0.0026&18.8&0.3        &3.        & 0.5      & 2.1       & 0.0009  & 90.     & 0.197 \\
\ cal2  & 130    &80    & 4.8        &22     &9.5 &0.1        &5.6       & 0.7      & 0.8       & 0.056   & 590      & 0.109 \\
\ cal3  & 70     &90    &4.3         &0.014  &8.  &0.3        &1.8       & 0.3      & 1.        & 0.0046  & 283     & 0.193 \\
\ cal4  & 70     &160   &4.2         &0.018  &2.4 &0.15       &2.        & 0.3      & 2.5       & 0.014   & 626     & 0.285\\
\ cal5  & 70     &170   &4.2         &0.018  &2.67&0.15       & 2.5      & 0.3      & 0.8       & 0.015   & 692     & 0.3989\\
\ cal6  & 80     &190   &4.2         &0.018  &1.98&0.15       & 2.       & 0.3      & 0.8       & 0.025   & 1000    & 0.4006\\
\ cal7  & 70     &170   &4.2         &0.018  &2.7 & 0.15      & 2.5      & 0.3      & 0.8       & 0.015   & 690     & 0.3996\\
\ cal8  & 80     &160   &4.3         &0.016  &1.8 & 0.1       & 1.5      & 0.1      & 0.15      & 0.012   & 800     & 0.3994\\
\ cal9  & 120    & 90   &5.5         &0.2    &8.  & 0.4       & 6.5      & 0.7      & 0.1       & 0.024   & 648     & 0.3996\\
\ cal10 & 120    & 90   &4.9         & 0.002 &8.  & 0.1       & 4.5      & 0.7      & 0.2       & 0.0023  & 647     & 0.3989\\
\ cal11 & 110    & 90   & 4.         &0.0001 & 8. & 0.1       & 4.       &0.4       & 0.2       & 0.00054 & 585     & 0.496 \\
\ cal12 & 60     & 50   & 4.4        & 0.015 & 8. & 0.1       & 4.5      & 0.2      & 0.1       & 0.0019  & 119     & 0.425\\
\ cal13 & 120    & 90   & 4.9        & 0.008  &8. & 0.1       & 3.       & 0.2      & 0.2       & 0.0058  & 643     & 0.4311\\
\ cal14 & 130    & 90   & 4.5        & 0.025  & 8.& 0.3      & 6.6      & 0.4      & 1.        & 0.012   & 700     & 0.2343\\ \hline        

\end{tabular}
\end{table*}

\begin{table*}
\caption{The sample of LINERs by Contini (1997)}
\begin{tabular}{lclclclclclclclclc} \hline  \hline
\ galaxy  & z      & galaxy  & z     &   galaxy  & z  &  galaxy  & z      & galaxy  & z     &   galaxy  & z  \\ \hline
\ NGC 315 &0.0165  & NGC 404 &0.0182 & NGC 1052  &0.005 &NGC 1167&0.0165  & NGC 1275&0.0175 & NGC 1667  &0.0139\\
\ NGC 2639&0.011   & NGC 2841&0.0024 & NGC 3031  &0.0001& NGC 3504&0.005   & NGC 3642&0.0053 & NGC 3998  &0.0035\\
\ NGC 4395&0.001   & NGC 4618&0.002  & NGC 7217  &0.0031& NGC 7479&0.008   & NGC 7714&0.0094 & NGC 7743  &0.0056\\ \hline
\end{tabular}

\end{table*} 

\begin{table*}
\caption{The sample of SB galaxies by Viegas et al. (1999)}
\begin{tabular}{lclclclclclclclclc} \hline  \hline
\ galaxy   &   z   &   galaxy  &   z   &   galaxy   &    z   & galaxy   &   z    \\ \hline
\ II Zw 40 & 0.0027& NGC 5253   & 0.0014& NGC 3690  & 0.01   & NGC 3256 & 0.0093 \\
\  M82     & 0.008 & M83        & 0.00165&NGC 253   & 0.0158 & NGC 4945 & 0.0019 \\ \hline
\end{tabular}
\end{table*}

\begin{table*}
\caption{Comparison of calculated with observed line ratios to \Hb for the  Winter et al (2010) sample}
\begin{tabular}{lccccccccccccccc} \hline  \hline
\ galaxy  & FWHM$^1$  & [OIII]5007+ & [OI]6300+ &[NII]6548+ & [SII]6717+ & z  \\ \hline
\ NGC 3516 & 582      & 9.28        & 0.345     & 3.0       & 1.2        & 0.009\\
\ m1       &   -      & 8.9         & 0.34      & 3.1       & 1.62       & - \\
\ 3C 105   & 240      & 17.07       & 0.33      & 2.31      & 1.17       & 0.089\\
\ m2       & -        & 17.12       & 0.446     & 2.88      & 0.8        & -   \\
\ 3C 111   & 214      & 14.01       & 0.3       & 0.57      & 0.81       & 0.049 \\
\ m3       & -        & 14.5        & 0.33      & 0.74      & 0.82       & -   \\
\ Mrk 110  & 483      & 7.98        & 0.49      & 0.86       & 0.7        & 0.035 \\
\ m4       & -        & 8.43        & 0.24      & 1.        & 0.74       &  -  \\
\ m4sb     & -        & 8.6         & 0.23      & 0.64      & 0.9        &  -  \\
\ Mrk 1498 & 320      & 4.21        & 0.15      & 0.78      & 0.75       & 0.055  \\
\ m5       & -        & 4.4         & 0.144     & 0.86      & 0.79       &  -    \\
\ m5sb     & -        & 4.1         & 0.2       & 0.85      & 0.86       & -     \\ 
\ 3C 403   & 270      & 11.62       & 0.48      & 2.85      & 1.77       & 0.059 \\
\ m6       & -        & 12.3        & 0.5       & 2.6       & 1.2        &  -   \\
\ Mrk 1018 & 693      & 9.91        & 0.33      & 4.68      & 2.08       & 0.043 \\
\ m7       &  -       & 10.         & 0.3       & 4.2       & 1.0        &  -  \\
\ Mrk 813  & 1486     & 4.74        & 0.21      & -         & 1.50       & 0.11 \\
\ m8       &  -       & 4.8         & 0.23      & 1.12      & 1.         & -    \\
\ m8sb     &  -       & 4.49        & 0.16      & 0.37      & 0.9        & - \\
\ NGC 4051 & 227      & 4.5         & 0.46      & 2.1       & 1.19       & 0.002\\
\ m9       &  -       & 4.4         & 0.4       & 2.6       & 1.0        & - \\
\ NGC 4151 & 488      & 11.56       & 0.75      & 2.3       & 1.84       & 0.003 \\
\ m10      & -        & 11.         & 0.9       & 2.8       & 1.2        &  - \\ \hline 

\end{tabular}

$^1$ in \kms

\end{table*}

\begin{table*}
\caption{The physical conditions for the AGN models in the Winter et al (2010) sample}
\begin{tabular}{lccccccccccccccc} \hline  \hline
\   &            z   &   \Vs & \n0 &    $F$ &   N/H & O/H        &  S/H      & $D$     & \Hb abs  \\       
\   &                &   \kms & \cm3 & units $^1$& 10$^{-5}$  &10$^{-4}$& 10$^{-5}$ & 10$^{17}$cm & units $^2$  \\  \hline
\ NGC 3516 &   0.009&  550 &  40 & 0.5 &  10.   &  6.6   &    1.8   &3.6 & 0.019 \\
\ 3C 105   &    0.089&  200 &  250&  10. &  7.    &  6.6   &   1.8   &1.19   & 0.185 \\
\ 3C 111  &    0.049&  200 &  250&  8.  &  2.    &  6.6   &    1.8   &1.1    & 0.18 \\
\ Mrk 110  &    0.035&  400 &  50 & 0.9 &  5.    &  6.6   &   1.5   &75.    & 0.043 \\
\ Mrk 1498 &   0.055&  320 &  400&  1.&  2.    &  5.6   &    2.5   &0.029 & 0.15 \\
\ 3C 403   &   0.059&  200 &  250&   8.&  7.    &  5.6   &    2.2   &15.4 & 0.3 \\
\ Mrk 1018 &    0.043&  700 &  350& 0.7 &  10.   &  6.6   &   2.9   &0.042  & 0.005\\
\ Mrk 813  &    0.11 &  1400&  70 & 0.8 &  2.    &  4.6   &   1.5   &0.123 & 0.02\\
\ NGC 4051 &    0.002&  250 &  230& 0.8 &  5.    &  6.6   &   1.5   &0.096  & 0.012\\
\ NGC 4151 &    0.003&  500 &  40 & 0.4 &  7.    &  6.6   &   1.5   &14.  & 0.012\\ \hline
\end{tabular}

$^1$ in 10$^{10}$  photons cm$^{-2}$ s${-1}$ eV$^{-1}$ at the Lyman limit;
$^2$ in \erg

\caption{The physical conditions for the  SB models in the Winter et al (2010) sample}
\begin{tabular}{lccccccccccccccc} \hline  \hline
\         &   z     & \Vs & \n0 & \Ts   &  $U$   &  N/H   & O/H    &  S/H    & $D$   & \Hb abs  \\       
\         &         & \kms& \cm3& 10$^4$ K& -    &  10$^{-5}$&  10$^{-4}$ & 10$^{-5}$ &10$^{17}$ cm & units $^1$ \\ \hline
\ Mrk 110 &   0.035  & 400  &  50  & 7.  & 1.     &  5. & 6.6& 0.5  & 20.   &  0.044 \\
\ Mrk 1498&   0.055  & 320  &  400 &  6. &0.1     &4.   & 6.6&  1.3 &10.   & 0.23  \\
\ Mrk 813 &   0.11   & 1400 &  100 &  6.5& 1.     &2.   & 4.6&  1.5& 10.  & 0.26 \\ \hline
\end{tabular}

$^1$ in \erg

\end{table*}

\subsection{Star-forming galaxies and compact narrow emission-line galaxies (0.11 $<$z$<$ 0.5)}

The spectra of HII regions in 14 star-forming emission-line galaxies (ELGs) at 0.11 $<$z$<$ 0.5
were  reported  by Kobulnicky \& Zaritsky (1999, hereafter KZ99), in order to investigate, in particular, the N/H and O/H
relative abundances. The spectra were obtained by the Keck II telescope and low-resolution imaging
multiobject spectrograph.
The  emission line widths of the observed galaxy sample are smaller than those of the local ones.
Moreover KZ99 claim that these galaxies are  slightly more
metal-deficient and that metallicity of a galaxy increases monotonically with age.

We have modelled these spectra by SB models  as we have done for  previous galaxies. 
The results are presented in Tables 5 , 6  and 7 .
We have found, by a detailed modelling, lower than solar relative abundances for N/H, O/H and Ne/H,
by factors $\leq$ 10, between $\sim$ 1 and 4  and $\leq$3,
respectively. S/H  shows lower than solar values even in local galaxies because  S can be easily trapped into dust grains.
KZ99 have found by direct O/H measurements that O/H ranges from a minimum of 10$^{-5}$ for SA 68-206985 to
a maximum of 1.9 10$^{-4}$ for  SDG 223, which are lower than solar by factors of 66 and 3.5, respectively.
They found  lower  N/O relative abundances by a factor $\sim$ 10.
The values obtained by the KZ99 empirical method are higher.

Table  4 shows that compared with the  KZ99 sample of star forming
galaxies, Capetti et al.  3C 258 relic galaxy corresponds to a shock velocity about double and a
similar preshock density.
 The star temperature is slightly higher and the ionization parameter is similarly low, as well as the N/H and S/H
relative abundances. On the other hand, O/H is about solar.

\subsection{LINERs and SB galaxies at z$\leq$0.03}

So far the distribution of the AGNs at low z
 along the redshift axis for both AGN and  SB is rather poor.
Therefore, we have  inserted in our sample  the physical parameters which
 were calculated for other galaxies in previous works, e.g. for LINERs from the sample (Table 8)
presented by Contini (1997) based on the observations of Ho
et al. (1993). Moreover, we have added the results from the sample of  SB galaxies  (Table 9) presented by
Viegas et al (1999). In Tables 8  and 9  the galaxy names are followed by their redshift.

Still, we could not cover the whole range of z between
0.001 and  0.11 uniformly.

\subsection{AGN spectra from  the Winter et al. (2010) local galaxy sample}

 We   collected the spectra of galaxies at the redshifts where the  calculated data were missing,
selecting a subsample of AGNs
from the $SWIFT$ Burst Alert Telescope survey, that  was reported  by  Winter et al. (2010).
Spectra including lines such as \Ha, \Hb, [OIII] 5007+, [OI] 6300+, [NII] 6584+, [SII] 6717+ are  suitable to a reliable modelling.
 We present the results of  modelling in Tables 10, 11 and 12.
 Some of the selected galaxies, e.g. NGC 4151,  were observed and modelled previously in different locations
throughout the galaxy (Contini, Viegas, Prieto  2002 and references therein). We chosed Winter et al
data because referring to  averaged spectra.
As we have done for the Ramos Almeida et al (2013) sample in Paper I, we selected some objects whose spectra  could  be  satisfactorily modelled
 also by  starburst models  showing  the eventual contribution of starbursts to the AGN.
In Table 10 the observed corrected data for each galaxy  are followed by the model results in the next rows.
Models m1-m10 refer to the AGN and the results are reported in Table 11. Models m4sb, m5sb and m8sb refer to the SB
and are reported in Table 12.  For all the models Ne/H =10$^{-4}$ is adopted.

\section{Evolution of the physical parameters and of  the relative abundances}

We  assume  as $results$ of our analysis the  input parameter set which leads to the best fit of
the  line ratios (and of the continuum SED, when observed) for each single object.
 The  modelling precision is shown in Fig. 2 where
the calculated and observed [OIII]5007+/\Hb and [NII] 6548+/\Hb line ratios are compared  (left
and right diagram, respectively).
 For  the samples  containing one only  galaxy  the  approximation of  model results
to the data can be easily estimated from  the tables, therefore they are omitted in Fig. 2. 
The observed [OIII]5007+/\Hb for Ramos Almeida  et al. (2013) galaxy G107
is an upper limit, therefore Fig. 2 shows that  this line ratio is not exactly reproduced by the model. 
The comparison of [NII]/\Hb line ratios  in  Fig. 2 right diagram shows some scattering.
In fact the observed [NII] lines can be affected by blending  with \Ha.
Fig. 2 shows that  Ramos Almeida et al. (2013) G60 galaxy is more likely AGN dominated.

\begin{figure*}
\includegraphics[width=8.6cm]{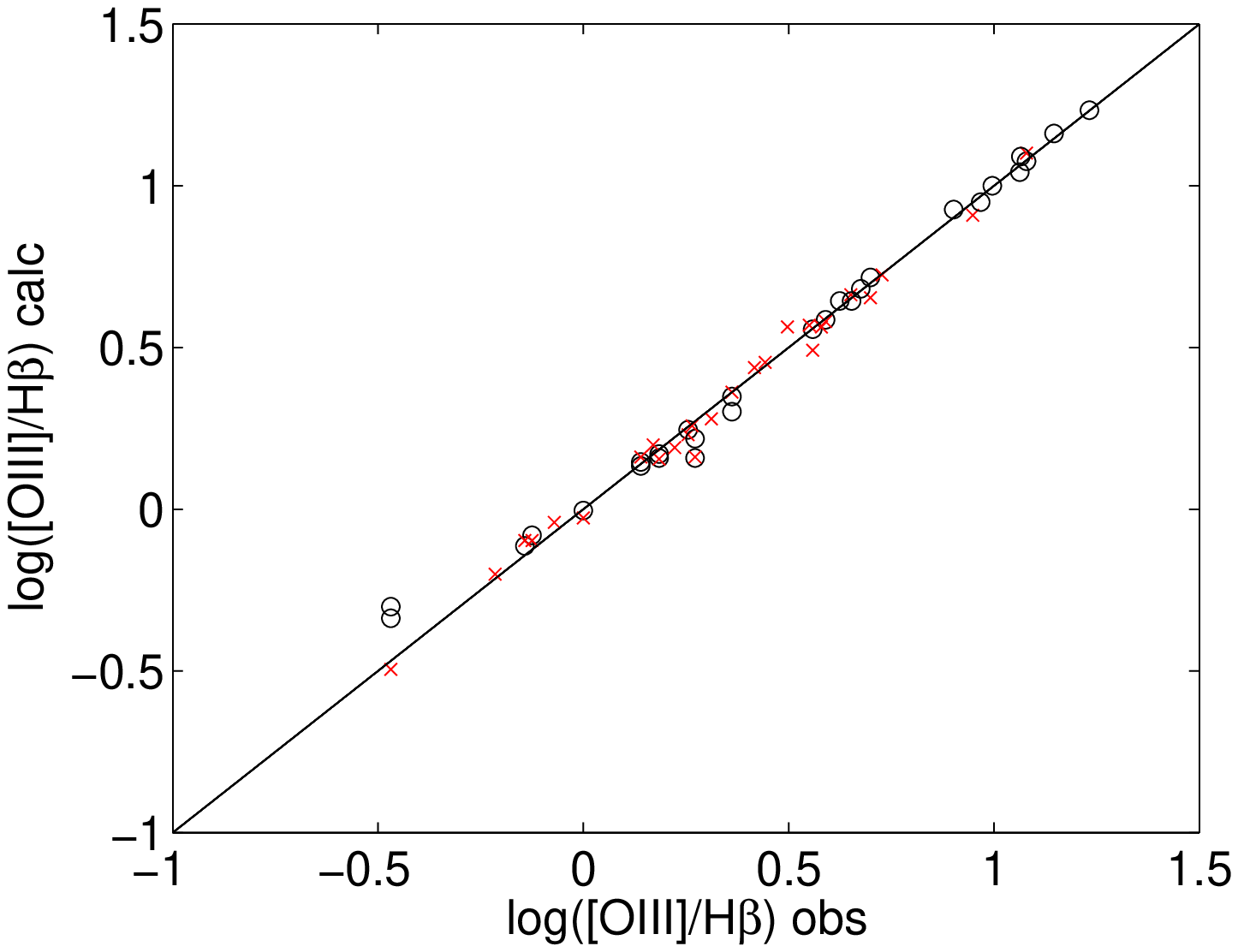}
\includegraphics[width=8.6cm]{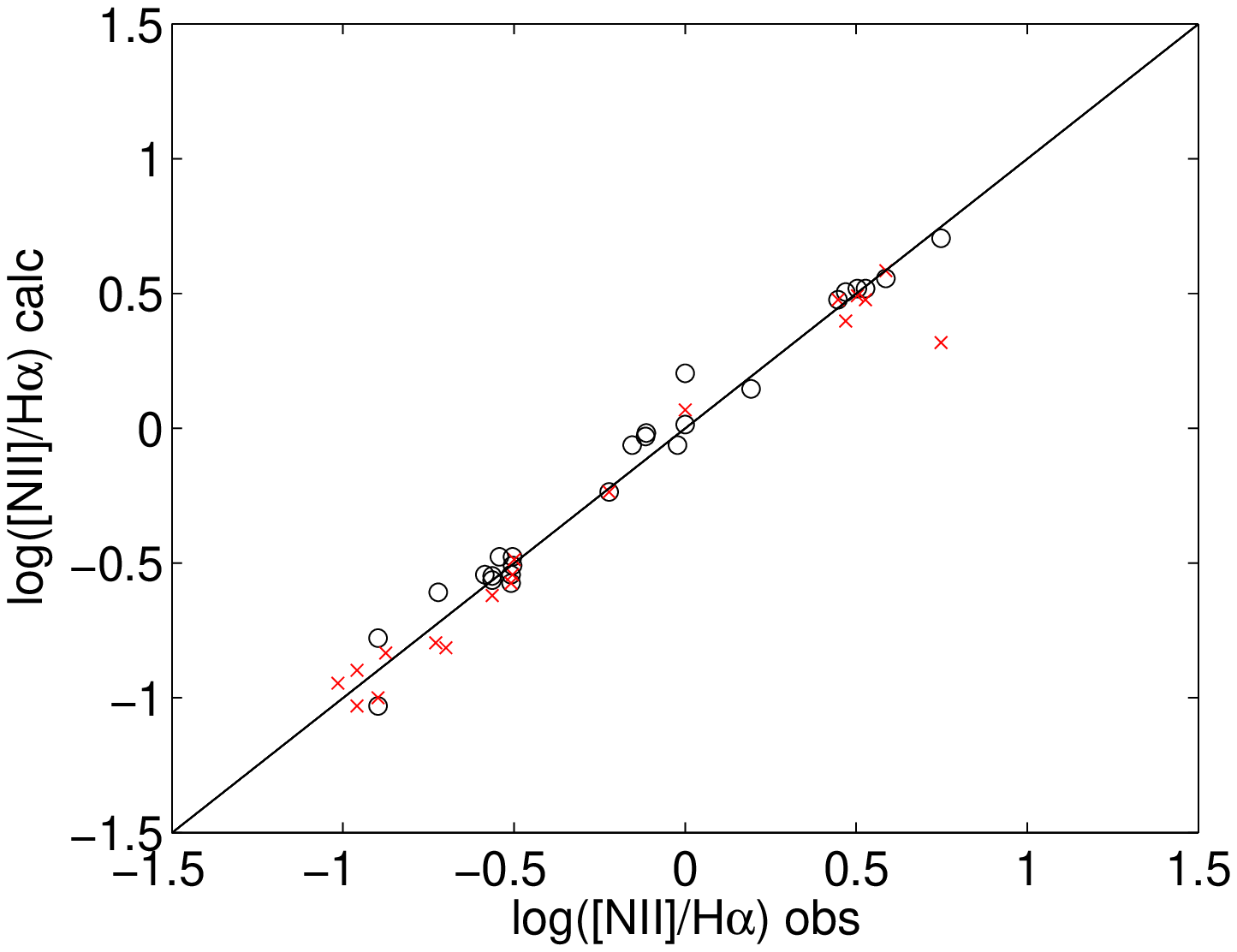}
\caption{Comparison between calculated and observed line ratios for the  galaxies.
Models for AGN : black circles;  SB models : red crosses; the solid line is the 1:1 line.}
\end{figure*}

The trends of the physical parameters and of the relative abundances calculated by modelling
AGN and SB spectra of galaxies at redshifts 0.001 $\leq$z$\leq$ 3.35 are presented in Fig. 3.
Besides \Vs, \n0, $F$, $U$, \Ts,  $D$ and the relative abundances, we  draw special  attention
to  the \Ha absolute flux \Haa, calculated for the single objects, and to the geometrical thickness $D$
of the emitting clouds which cannot be predicted directly from the line ratios but only from  model calculations.
They are connected with the luminosity of the galaxies and with  the emitting  cloud dimensions,
respectively. \La   measured at Earth is related  with the  \Ha flux emitted by the nebulae (the NLR for AGNs and the
region surrounding the SB) throughout the galaxy and with
  the radius of the NLR and SB regions $R$ :
%
\La (Earth) = \Ha(Earth)$\times 4 \pi d^2$ = \Haa(nebula)$\times 4 \pi R^2$ , where  d is the distance to Earth.

\subsection{Parameters referring to  the shock}

 Fig. 3  top left panel refers to \n0 and \Vs.
Preshock densities in the emitting clouds range between $\sim$ 50 and $>$ 1300 \cm3.
Recall that the density in the downstream
region of the clouds, where both the line and continuum spectra are emitted,
is higher than \n0 by a factor of  4  to $>$100, depending on the shock velocity.

We  have found that
galaxies at  0.4$\leq$z$\leq$  3  show  relatively high preshock densities (500-1300 \cm3) which may indicate that the gas
has been compressed by shock waves created by  collision during galaxy merging. 

The density increases with  decreasing z from z=3.35 to z=0.5 which corresponds to the top of a broad \n0 peak. 
Then, \n0 decreases towards z=0.1. At low z, \n0 regains values between $\sim$100 and 600 \cm3,  the highest for SB galaxies.

Shock velocities  were  derived as a first choice from  the  FWHM  of
line profiles of  each object referring to the NLR or to the SB surrounding clouds. 
Then they were slightly  modified cross-checking  
the modelling of  the line ratios. \Vs ranges between
100 and 1600 \kms,  with  an increasing trend with z from z=0.6 to z=3.35.
The opposite trends of \Vs and \n0 in the same z range are in agreement with the
Rankine-Hugoniot law of conservation of matter throughout the shock front and downstream.
 This suggests that the dynamical parameters, e.g. \Vs and \n0 at relatively low z,
are  most probably connected with phenomena on  small scales  in  regions within  the galaxy, 
rather than with evolution
of galaxies on a Hubble time scale.

The shock velocities deduced from modelling  are  in
the same range  for AGN at high and low z. The lowest \Vs appear for the HII star forming regions
of compact galaxies (KZ99). 

\subsection{The photoionising radiation flux}

Fig. 3  top right panel shows the trend of the temperature of the stars \Ts
(in units of 10$^4$ K)  in SB galaxies, in the SBs included in the 
LINER sample and in  samples at higher z. 
The stellar temperature ranges between 10$^4$ K and $\geq$ 10$^5$ K.  
A  decreasing trend of \Ts with increasing z can be noticed. Interestingly, the
highest \Ts   are found in SB LINERs and the lowest ones  in the starbursts calculated for
optically faint  ULRIGs at relatively high z.
This indicates that  the  stars are younger in local galaxies. 
The stars  formed in SB at higher z  had  the time to cool down.
Some stars in the LINER sample galaxies which show SB characteristics, show temperatures of $\sim$ 10$^5$ K,
which are suitable to Wolf-Rayet stars. 
In fact they appear in the z range  corresponding to higher than solar N/H.
W-R stars were recently predicted in merger galaxies (Contini 2012a).
The outstanding \Ts in galaxies at z$\sim$ 2 are discussed in Contini (2013, inpreparation).

The flux from the AGN $F$  (in 
units of 10$^{10}$ photons cm$^{-2}$ s$^{-1}$ eV$^{-1}$ at the Lyman limit)
and the ionization parameter $U$ appear together in the diagram. 
The ionization parameter for LINERs is low,  even lower than $U$ calculated for compact galaxies (KZ99)
at higher z.
This could indicate  that the flux is diluted by the  large geometrical thickness $D$ 
of the emitting clouds which  can reach 1-10 pc in LINERs (Contini 1997).
Alternatively, the radiation flux is  obstructed on its way to the emitting clouds by patches of
 gas and/or dust.
For the  starbursts which could contribute to the AGN spectra observed by Ramos Almeida et al at higher z,
$U$ increases by a factor of $\leq$ 100. We wonder how models calculated by these high $U$ could
fit  eventually observed  IR line ratios (cf Contini 2013a).
Both $F$ and $U$  increase  with z decreasing  from z=3.35 to z=0.5. The shock velocity  decreases in this range of z
and the preshock density in SB clouds are low. 
Therefore the flux is less absorbed. Moreover, $D$ has not yet  reached the  large thickness as those found 
for galaxies at z$\leq$ 0.1.

Log $F$, which refers to AGNs in general,  shows a rough 
maximum (log $F$ $\geq$ 11) at z $\sim$ 0.1, 
decreasing towards lower redshifts.
The flux from the AGN in the Ramos Almeida et al (2013)  (Paper I, table 3 ) and Brand et al.  (Table 1) sample galaxies is moderately high 
($F$ = 5.10$^9$-2 10$^{10}$ photon cm$^{-2}$ s$^{-1}$ eV$^{-1}$
at the Lyman limit)) similar to that characteristic of low-luminosity AGN, 
while it ranges
within  more than three orders of magnitude  for the ensemble of the local AGNs.

\subsection{Relative abundances}
 
Fig. 3  left middle panel shows  O/H, N/H, Ne/H and S/H (in 10$^{-4}$ units) calculated consistently 
with the physical parameters by modelling the  spectra.
Unfortunately the spectra from Ramos Almeida et al (2013) reported in Paper I and those of Brand et al. (2007) 
do not contain  data for the Ne and S lines.

The Ne/H relative abundances are solar  and higher than solar (10$^{-4}$ Allen 1976, 1.3 10$^{-4}$ Cunha et al 2006) 
up to z$\sim$ 0.1
and  slightly decrease at higher z. The S/H relative abundances are solar
at low z. They  slightly decrease up to z$\sim$ 0.4 where they abruptly drop by a factor 
of $\sim$ 10 for KZ99 star-forming HII regions. Sulphur  can be easily trapped into dust grains.
This trend is  similar to that of  N/H and O/H for KZ99 objects in agreement with KZ99 prediction
that these objects are metal poor.
The low metallicities in galaxies showing collision evidence are  generally explained by  
acquisition of external gas in  merging processes.
However, the
 O/H and N/H dips characteristic of KZ99 objects at z between 0.1 and 0.5 seem connected to the characteristic low
\Vs of the sample.
Lower than solar Mg/H and Si/H are also predicted for these objects.
At shock velocities $\leq$ 100 \kms collisional processes are less efficient, in particular, 
dust grain sputtering at the shockfront. Therefore O, N and S  remain trapped into  grains and  molecules.

The N/H relative abundances for AGNs  show  fluctuations  with  a maximum   almost twice solar at z$\leq$ 0.01
and a minimum  $\leq$ 0.5 solar at z$\sim$ 0.1.
In the starburst clouds  N/H  drops to  lower than solar values by a factor  $\leq$3 at
 z$\sim$ 0.1.   
 The decrease  of N/H and O/H from solar to $\sim$ 0.1 solar with z decreasing from 3.35 to 1 may
indicate that  at this redshift range dust and stars are forming, reaching their maximum rates at different z between
z$\sim$1 and z=$\sim$0.16.

O/H is nearly solar (6.6 10$^{-4}$) for most of the galaxies. Guaita et al (2013) calculated for star-forming 
galaxies at 2 $<$z $<$ 3
log(O/H)+12 = 8.2-8.8 (corresponding to O/H=1.58 - 8.3 10$^{-4}$). The  higher limit is in agreement with our results
for the SBs in the optically faint ULIRGs by Brand et al (2007).  
 The peculiar minimum of O/H and N/H ratios   calculated in Paper I  by  AGN models   
for  a few Ramos Almeida et al.  objects, G53 and G107  at relatively high z,
   confirm that these  galaxies are  SB dominated rather than AGN, as predicted by their location in the 
BPT diagram (Fig. 1). 

 The scattering of the relative abundance results in  the 0.1$\leq$ z$\leq$1. range  is evident also for 
the physical conditions
in nearly all the diagrams of Fig. 3, due to the different  types of galaxies involved. 
In particular, the geometrical thickness of the emitting clouds
increases by a factor $>$ 10 towards local galaxies in agreement with lower shock velocities
and a reduced fragmentation of matter.
 
 The results presented  in Fig. 3 show  in general  O/H higher than those calculated by
the direct methods.
This can be understood considering that using composite models
 the line intensities are calculated by  the fractional abundance of the ion  
corresponding to the line, e.g. O$^{++}$/O for [OIII],  (and by other factors such  as the collisional strength
  and the statistical weight)
which  are integrated throughout the cloud  recombination zone downstream.
Here the electron temperature and density of the emitting gas  
are not constant but follow the gas cooling rate.
The regions  corresponding to low  O$^{++}$/O are also accounted for.
Namely,

\noindent
  [OIII]/\Hb $\propto$ ($\Sigma$$_i$ (O$^{++}$/O)$_i$/(H$^+$/H)$_i$) [O/H] , where $i$ refers to the slab
downstream. The  O/H
 relative abundance  is  adjusted in order to  reproduce the observed line ratio.  
Using  the direct method,  the electron temperature  is  derived from the observed  
 [OIII]5007+/[OIII]4363 and the electron density from[OII]3727/[OII]3729  (the $\lambda$3727 and 
$\lambda$3729 lines are generally  blended). Then, the [OIII]/\Hb line ratio is calculated
adopting this temperature and this density and the O/H abundance is evaluated by
 fitting the observed line ratio.

\subsection{\Ha absolute flux and the geometrical thickness of the emitting clouds}

The  middle right panel of Fig. 3  shows the trend of  \Haa, the absolute \Ha flux calculated  from the 
emitting clouds in the galaxy
 and  the geometrical thickness of the emitting clouds $D$.  

As already mentioned, $D$
in both  the SBs and the AGNs decreases strongly with increasing z from a maximum average of $\geq$ 1 pc
to $\sim$ 10$^{-3}$ pc. The clouds are  more extended in  SBs than
in AGN NLR. 
The geometrical thickness of the clouds corrsponding to the  LINER sample were not  specified for single objects 
presented by Contini (1997),
nevertheless  a   1 - 10 pc range is globally indicated,
enhancing even more the decreasing trend of $D$ with increasing z.  
The geometrical thickness of the emitting clouds ranges
within three magnitude orders. A rather steep drop  of $D$ can be noticed for   0.1 $\leq$z$\leq$ 1.

\Haa, following the trend of the ionization radiation flux $F$ along the redshift axis,
 shows a slightly increasing trend up to z=0.1 
and a decreasing trend  for z$\geq$ 0.3 for  AGNs. 
The SBs show some discontinuity, as already mentioned.

The fluctuations   for 0.1 $<$ z$<$ 0.5 of \n0 and \Vs calculated for the KZ99 sample
translate to fluctuations of \Haa, because \Haa is $\propto$ n$^2$. 

\subsection{The radius of the NLR and of the SB surrounding clouds} 

The  \Ha luminosities   were  collected from the literature or calculated by the observed \Ha fluxes
L$_{H\alpha}$ = 4 $\pi$ \Ha d$^2$, where \Ha is
 the reddening corrected  observed \Ha fluxes at Earth
 and d the distance of the galaxies 
from Earth  (calculated by the redshift, adopting H$_o$=73 \kms Mpc$^{-1}$).

The radius R of the emitting regions (the NLR for AGNs and the  emitting regions surrounding the  SB)
is calculated by   L$_{H\alpha}$ = 4 $\pi$ \Haa R$^2$ assuming   uniform conditions. 
\Haa is calculated at the emitting gas  nebula by the \Hb absolute flux which  appears  
in Tables  2, 3, 4,  7, 11 and 12 
for each model. L$_{H\alpha}$ and  R are shown  in the bottom left panel of Fig. 3.

Comparing the range of $D$ with the range of R,
a  factor of $\sim$  10$^2$ - 10$^3$ is found, indicating that many clouds are located in the NLR
and in the  SB surroundings up to a maximum distance from the AGN and from the stars, respectively, given by R.
 Fig. 3 shows that R results are scattered throughout more than two orders of magnitude for 0.3 $<$ z $<$ 1.

\begin{figure*}
\includegraphics[width=8.6cm]{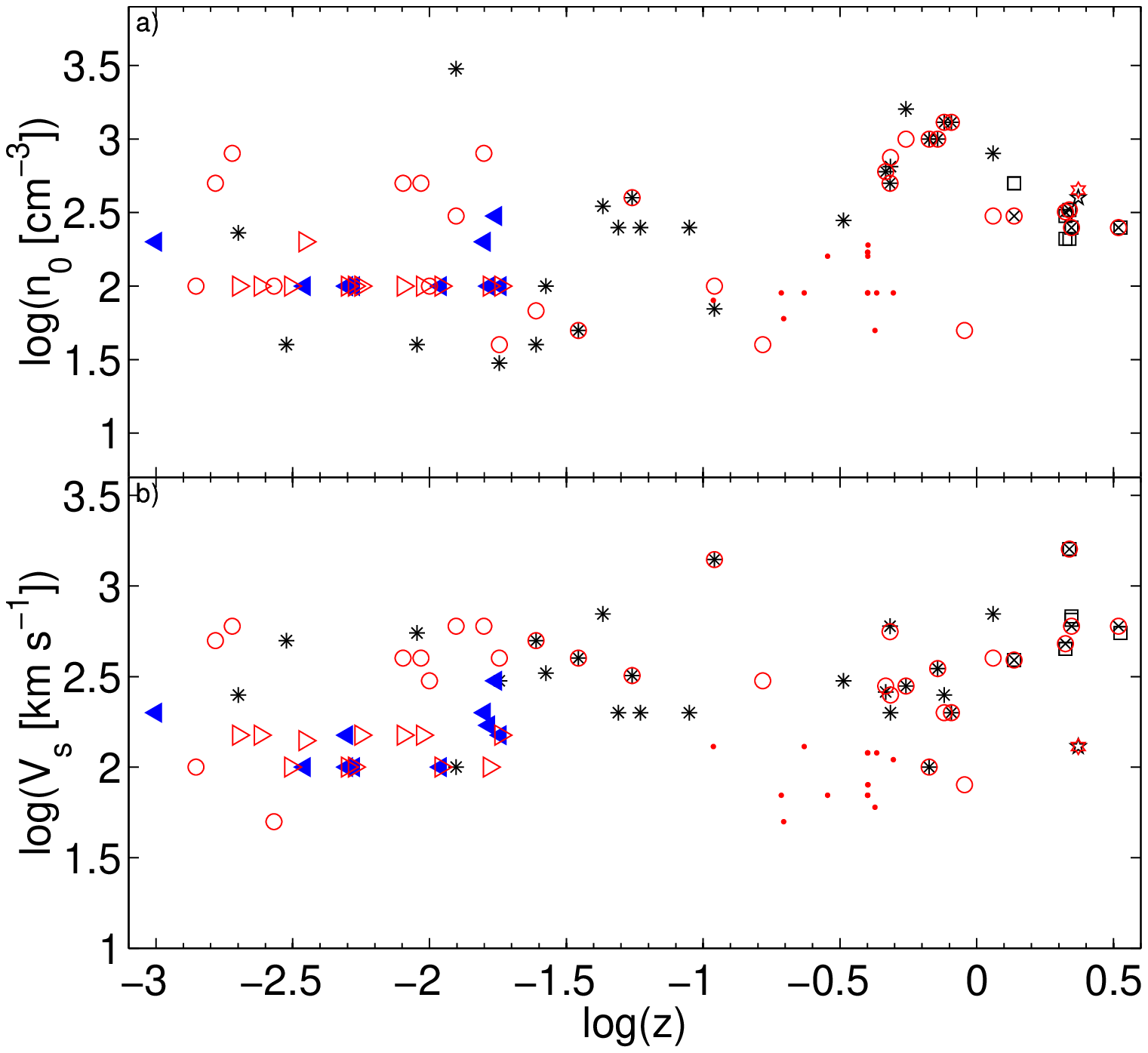}
\includegraphics[width=8.6cm]{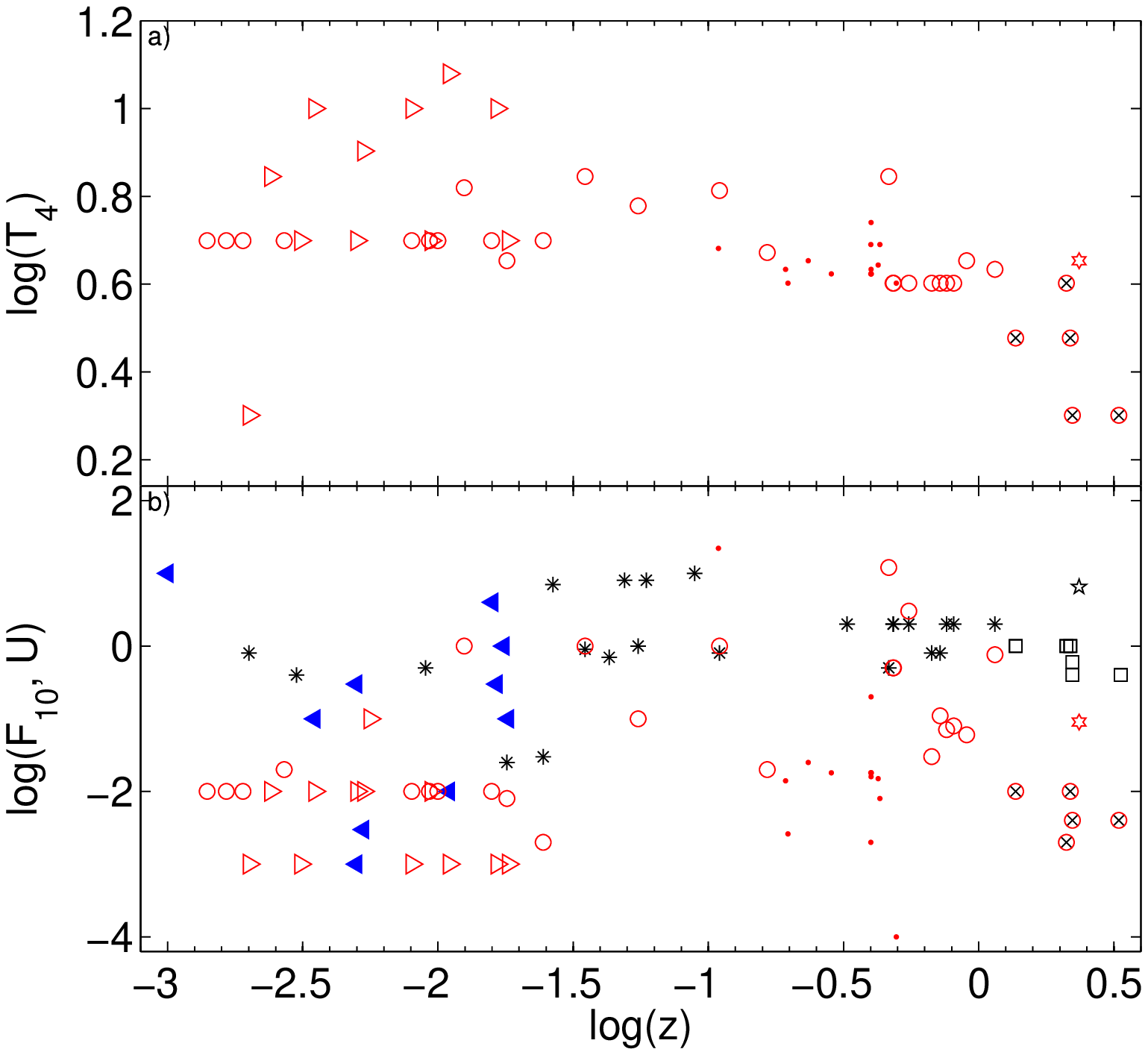}
\includegraphics[width=8.6cm]{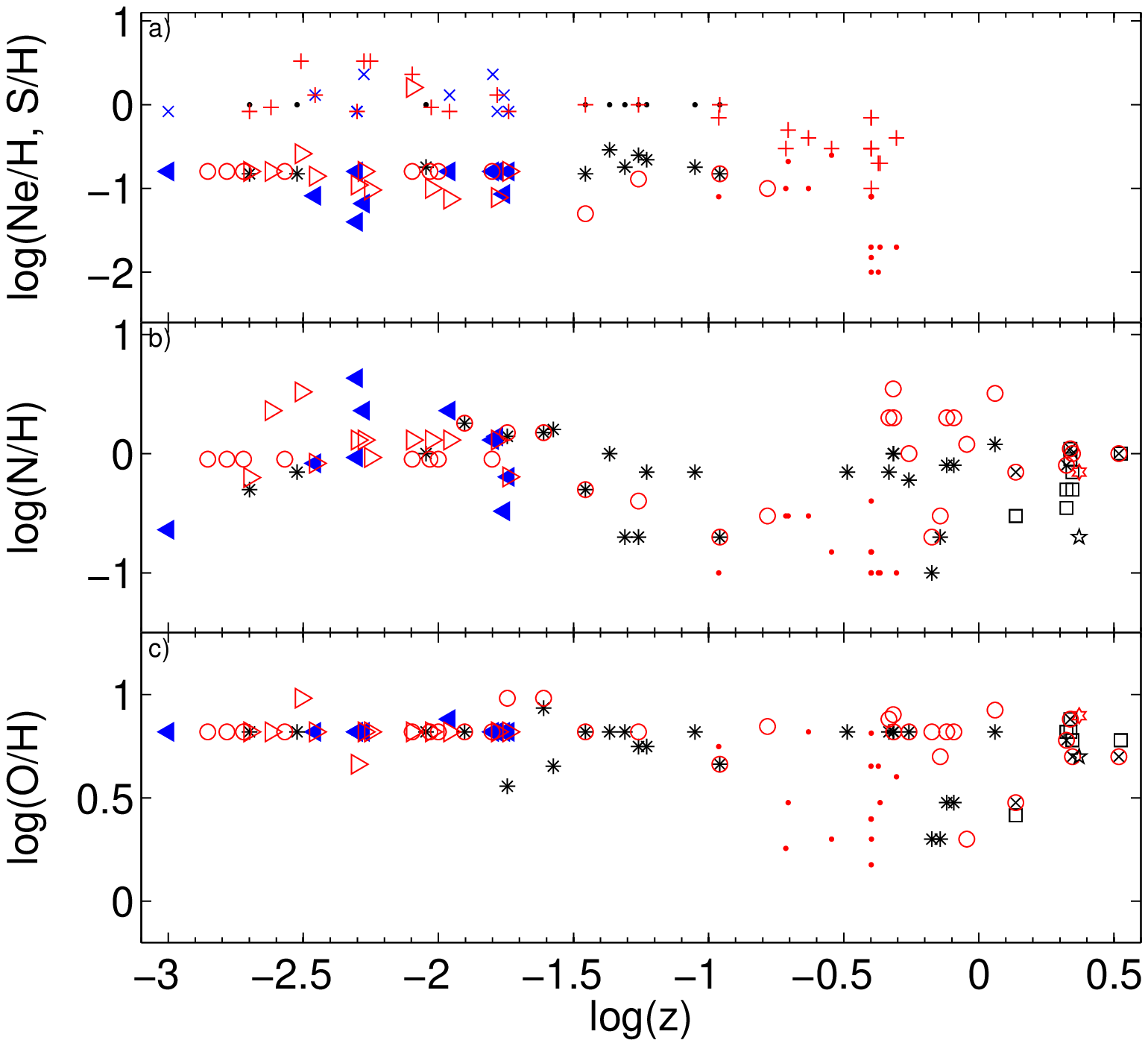}
\includegraphics[width=8.6cm]{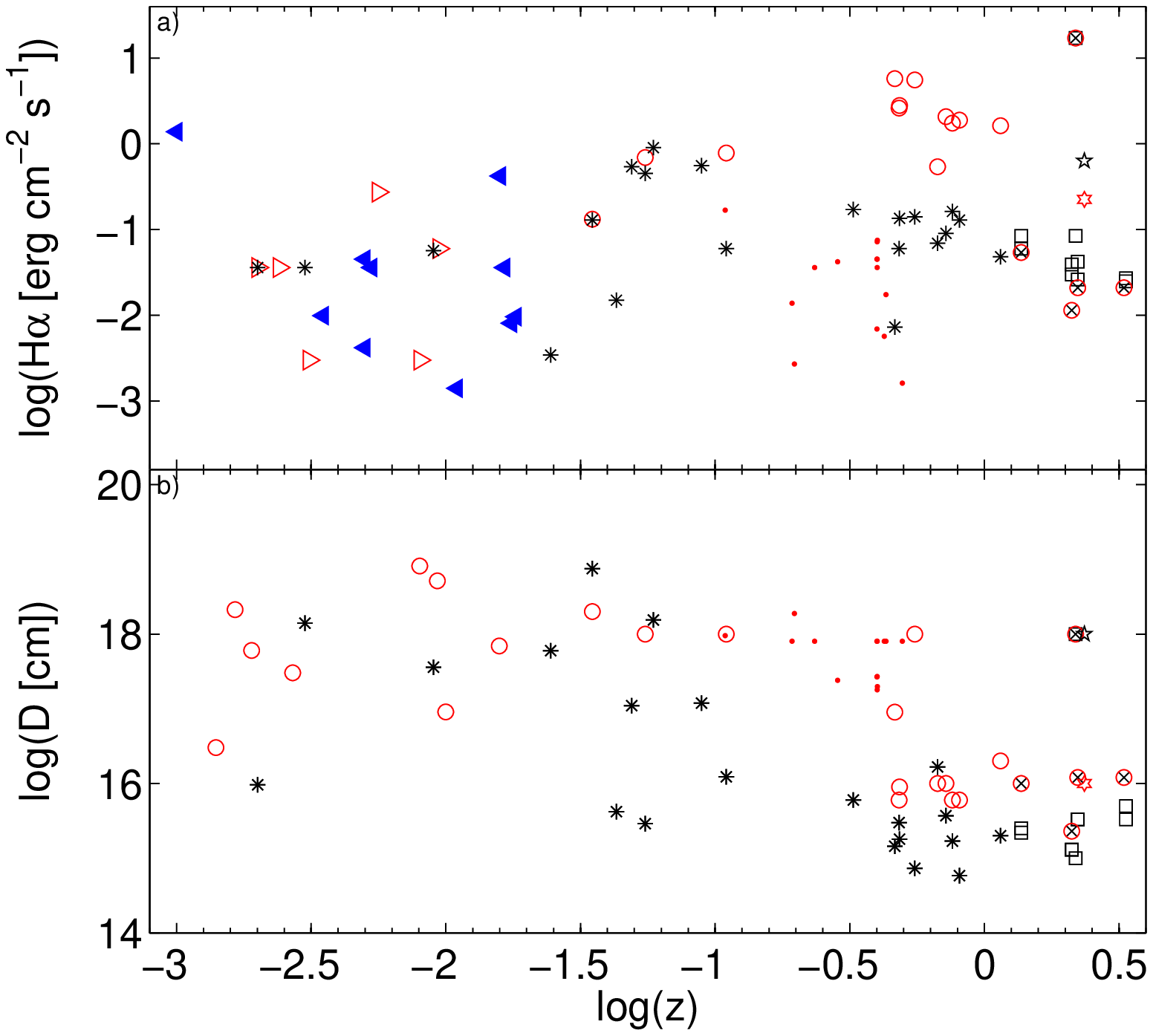}
\includegraphics[width=8.6cm]{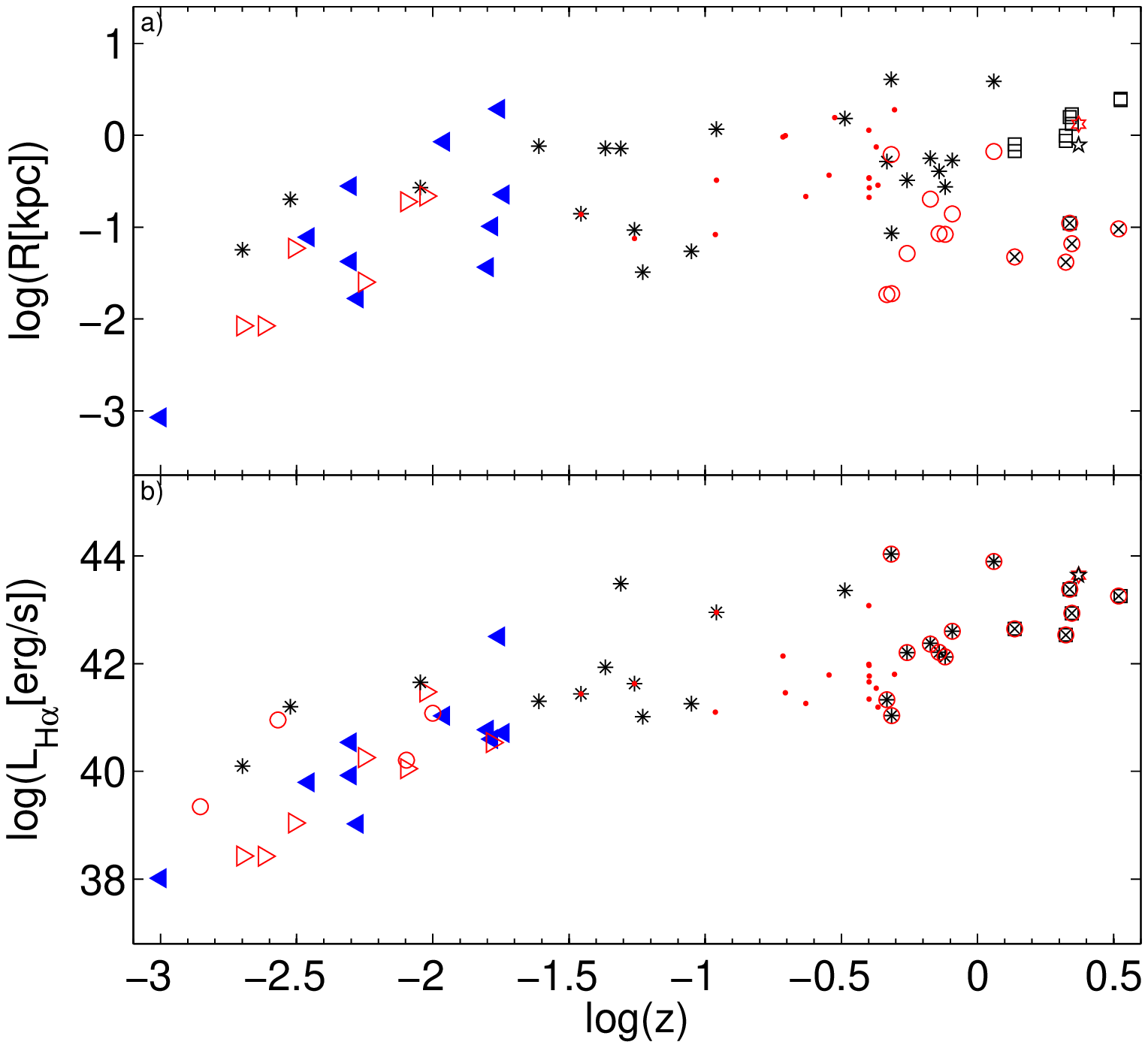}
\includegraphics[width=8.6cm]{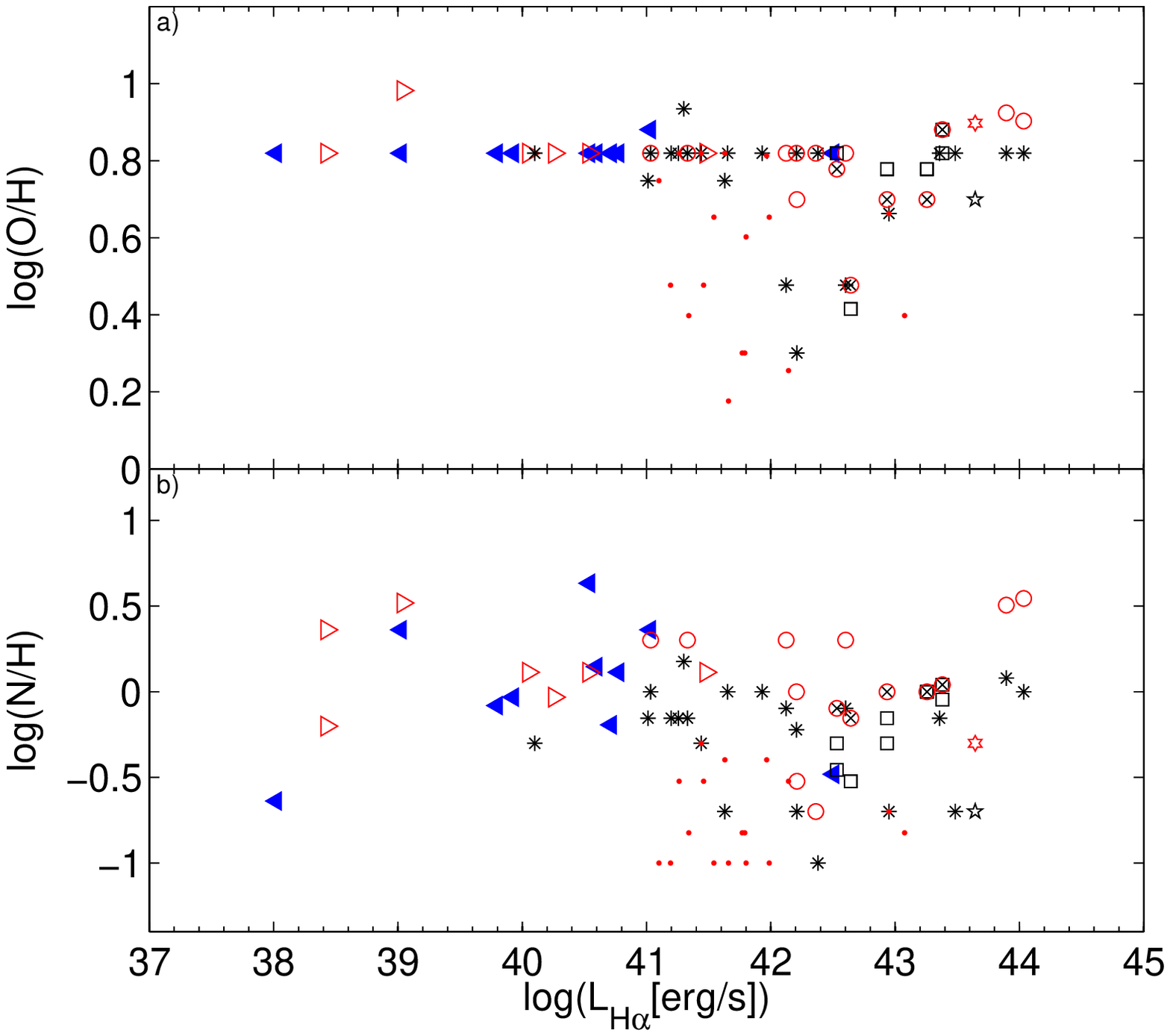}
\caption{
In all the diagrams red circles represent SB galaxies, red circles enclosing a x refer to the SBs in the optically faint ULIRGs;
red  triangles SB galaxies in the LINER
sample, red dots the KZ99 galaxies, red hexagram ULIRGS; black asterisks  the AGNs, black squares refer to the optically faint ULIRGs  and blue filled triangles the AGNs belonging to the
LINER sample; black pentagram ULIRGs. For Ne/H : red plus : starbursts; blue cross : LINER AGN ; black dots : AGN.
Top left : \n0 and \Vs; top right : the radiation parameters $F$ (in units of 10$^{10}$ phot cm$^{-2}$ s$^{-1}$ eV$^{-1}$ at
the Lyman limit), the ionization parameter $U$ and the temperature of the stars (in units of 10$^4$ K);
middle left : the relative abundances in units of 10$^{-4}$; middle right : \Haa and $D$; bottom  left:
the radius R of the NLR  in AGN  and of the emission regions in SBs
and the \Ha luminosity (L$_{H\alpha}$): bottom right : the metallicity versus L$_{H\alpha}$.
}
\end{figure*}

Concluding,
considering  the AGN feedback-driven star formation process (Ishibashi et al 2013), we have investigated 
the contribution of SBs
and  AGN to the galaxy  spectra. We  have found     star temperatures \Ts  $\sim$ 4 10$^4$ K at relatively low z.
The temperature of the stars decrease the higher the redshift, but a decreasing $U$ and an increasing R  hardly confirm  
compactness of galaxies at higher  z.
After some metallicity modulations in terms of higher than solar N/H at lower z and scattering of O/H between
solar and 0.2 solar   at 0.1$<$z$<$1, the metallicities regain the solar values at z $\sim$ 3.
The \Haa absolute fluxes emitted by the  SB clouds follow 
the trend of the ionization parameter and of the stellar temperature, while for AGN \Haa follows in particular 
the density trend.

\subsection{Star formation rates}

We   now consider  the  SFR trend 
 both for AGN and  SB galaxies along the redshift axis.
Generally the SFR  is calculated by the luminosities in the near UV for high z galaxies
and in the  FIR when the data in the IR  are available.
Actually,  we have not enough data to calculate  L$_{IR}$ for each object of our sample, but
we can   evaluate  the \Ha luminosity.

The  efficiency in forming stars is due to the number of photons  reaching the gas and dust clouds  from the
photoionizing source and 
to compression by the shock   in the emission line zone throughout the galaxy, i.e. the NLR in AGNs and  
 the regions surrounding the SB.

Fig. 3 bottom left panel  indicates that  L$_{H\alpha}$   increases with z.
If  the ionizing radiation comes from stars with mass  M $>$ 10$^{10}$\msol ,
 SFR(\msol yr$^{-1}$) = 7.9 10$^{-42}$  L$_{H\alpha}$ (erg s$^{-1}$).
So the SFR  decreased slowly  from  $>$ 100 \msol yr$^{-1}$ at z$>$1.15 for AGNs to
  $\leq$  0.001  \msol yr$^{-1}$ for  low luminosity LINERs at z $\leq$0.001,
  with a  roughly smooth  trend.

The metallicity - redshift relation for massive stars is discussed by Pilyugin et al (2013) considering
the various methods adopted to determine the O/H ratios from the spectral lines.
They suggest that there is no correlation between oxygen abundance and specific star formation rates
 in massive galaxies.
 Neglecting the SFR - mass relationship (see e.g. Mannucci et al 2010)
we present in Fig. 3 bottom right panel log (N/H) and log (O/H) versus  log (\La)
for a  small number of galaxies, compared to the sample adopted by
Pilyugin et al (2013 and references therein).  
Our results show that 
metallicity (in terms of O/H)
correlates with  \La (and therefore with SFR) particularly for SB  galaxies with \La $\geq$ 10$^{41}$ erg s$^{-1}$
 A rough correlation is seen also for N/H in the same range of \La with a large scattering due to the
different types of galaxies. 
There seems to be no correlation for AGN and for  low luminosity SB galaxies 
because O/H is roughly   constant  at the solar value. The critical redshift is $\sim$ 0.1$\leq$z$\leq$ 1.

\section{Concluding remarks}

We have  modelled an heterogeneous sample of   AGN and SB spectra observed  at intermediate-high and low redshifts, in order to
explore the evolutionary trend of some significant parameters.
The sample is relatively poor in number of objects, but  each of them has been  analysed  in details.

 Our investigation started by modelling the  spectra  presented by Ramos Almeida et al (2013)  for
 galaxies at intermediate redshift (0.27$\leq$z$\leq$1.28) in Paper I and ULIRGs at z between 1.37 and 3.35.
 Then  we compared their physical conditions with those of  galaxies at lower  and higher z.
Our approach  was the following.
After  selecting the observed spectra  containing enough number of lines  to avoid degeneracy,
the models
were constrained by the   comparison of  calculated  with  observed  line ratios.
We adopted  composite models  which account consistently for the flux
from a radiation source, outside  the emitting clouds,
 combined with the shocks accompanying the gaseous cloud motion.

At relatively high redshift, the data become rare.  
The results show  (Fig. 3)  the evolution of
physical conditions and the relative abundances   in the gas within the galaxy sample
 on a large z range.
The  parameter trends  are  relatively smooth but  their slopes change  at different redshifts.
This  indicates that various process rates, such as e.g.   star birth,
 relaxation from compact to  diluted gaseous clouds, etc. have different
rythms.

Summarizing, the evolution picture between z=3.35 and z=0.001 which results from our calculations shows that:
1) the gas in the emitting  clouds  of both AGNs and SB galaxies had  a density broad peak at z$\sim$0.5, 
becaming less dense and more elongated  with time.
2) The flux from the AGN had a maximum at z$\sim$ 0.1. 
3) The star temperatures in SBs increased with time,
indicating younger stars in local galaxies. 
4) The  O/H relative abundances  split into various branches :  from solar  to 1/3 solar
 from z=3.35 to z=0.6.  N/H  shows a modulated trend.
5) The NLR of AGNs and the starburst surrounding regions  were more fragmented  
at  z $\geq$0.35 due to higher shock velocities.
6) Metallicity (in terms of O/H)
correlates with  \La (and therefore with SFR) for  massive galaxies with different trends for SBs and
AGNs for \La $\geq$ 10$^{41}$ erg/s. A rough correlation is seen also for N/H. 
There seems to be no correlation for low luminosity galaxies.

\section*{References}

\def\ref{\par\noindent\hangindent 18pt}
\ref Allen, C.W. 1976 Astrophysical Quantities, London: Athlone (3rd edition)
\ref Anders E., Grevesse N. 1989, Geochim. Cosmochim. Acta, 53, 197
\ref Baldwin J. A., Phillips M. M., Terlevich R.  PASP, 1981, 93, 5
\ref Brand, K., et al. 2007, ApJ, 663, 204
\ref Brinchmann, J., Charlot, S., White, S.D.M., Tremonti, C. Kauffmann, G.
, Heckman, T., Bronkmann, J. 2004, MNRAS, 351, 1151
\ref Capetti, A., Robinson, A., Baldi, R. D., Buttiglione, S., Axon, D.J.,
 Celotti, A., Chiaberge, M.	2013arXiv1301.5757C	
\ref Caputi, K. et al. 2007, ApJ, 660, 97
\ref Cardelli, J.A., Clayton, G.C., Mathis, J.S.  1989, ApJ, 345, 245
\ref Chung, J.,Rey, S-C., Sung, E-C., Yeom, B-S., Humphrey, A., Yi, W., Hyeong, J. 2013 ApJ, 767, L15
\ref Contini, M. 2013a, MNRAS, 429, 242
\ref Contini, M. 2013b, A\&A, submitted
\ref Contini, M. 2013c, MNRAS, arXiv1310.3619 (Paper I)
\ref Contini, M. 2012a, MNRAS, 426, 719
\ref Contini, M. 2012b, MNRAS, 425,120
\ref Contini, M. 2009, MNRAS, 399, 1175
\ref Contini, M. 1997, A\&A, 323, 71
\ref Contini, M., Aldrovandi, S.M. 1983, A\&A, 127, 15
\ref Contini, M., Contini, T. 2003 , MNRAS, 342, 299
\ref Contini, M., Cracco, V., Ciroi, S., La Mura, G. 2012, A\&A, 545, 72 
\ref Contini, M., Viegas, S.M., Prieto, M.A. 2002, A\&A, 399, 414
\ref Cunha, K., Hubeny, I., Lanz, T. 2006, ApJ, 647, L143
\ref Diaz A. I., Prieto M. A., Wamsteker W.  , 1988, A\&A, 195, 53 
\ref Edmunds, M.G. \& Pagel, B.E.J.  1978 MNRAS, 185, 77
\ref Fabbiano, G.; Wang, Junfeng; Elvis, M.; Risaliti, G.  2011, Natur, 477, 431	
\ref Fosbury R. A. E., Wall J. V.  , 1979, MNRAS, 189, 79 
\ref Fried, J. W.; Schulz, H.  1983, A\&A, 118, 166	
\ref Guaita, L., Francke, H., Gawiser, E., Bauer, F.E., Hayes, M., \"{O}stlin, G., Padilla, N. 2013, arXiv:1301.5600
\ref Gunawardhana, M.L.P. et al. 2013arXiv1305.5308G
2011, A\&A, 529, 149
\ref Ho, L.C., Filippenko, A.V., Sargent, W.L.W. 1993, ApJ, 417, 63
\ref Hodapp, K.W., et al. 2003, PASP, 115, 1388
\ref Ishibashi, W., Fabian, A.C., Canning, R.E. 2013, arXiv:1302.4998
\ref Izotov, Y. I.; Stasińska, G.; Meynet, G.; Guseva, N. G.; Thuan, T. X.  2006, A\&A, 448, 955
\ref Izotov, Y. I.; Guseva, N. G.; Fricke, K. J.; Stasińska, G.; Henkel, C.; Papaderos, P.	
2010, A\&A, 517, 90
\ref Izotov, Y.I., Guzeva, N.G., Thuan, T.X. 2011, ApJ, 728, 161
\ref Jannuzi, B.T.,  Dey, A.  1999, in ASP Conf Ser.191, Photometric Redshifts and the Detection of High Redshift Galaxies,
ed. R.Weymann et al (San Francisco: ADP), 111
\ref Kauffmann, G. et al. 2003, MNRAS, 346, 1055
\ref Kakazu, Y., Cowie, L.L., Hu, E.M. 2007, ApJ, 668, 853
\ref Kewley, L.J., Dopita, M.A., Sutherland, R.S., Heisler, C.A., Trevena, J. 2001, ApJ, 556, 121 
\ref Kobulnicky, H., Zaritsky, D. 1999, ApJ, 511, 118
\ref Krogager, J-K. et al. 2013, arXiv:1304.4231
\ref Madau, P., Ferguson, H.C., Dickinson, M.E., Giovalisco, M., Steidel, C.C., Fruchter, A. 1996, MNRAS, 283, 1388
\ref Mannucci, F., Cresci, G., Maiolino, R., Marconi, A., Gnerucci, A., 2010, MNRAS, 408, 2115
\ref McLean, I.S. et al. 1998, Proc. SPIE, 3354,566
\ref Osterbrock, D.E.  Astrophysics of Gaseous Nebulae and Active Galactic Nuclei. Mill Valley, CA: 
University Science Books; 1989.
\ref Pagel, B.E.J., Simonson, E.A., Terlevich, R.J., Edmunds, M.G. 1992, MNRAS, 255, 325
\ref Pilyugin, L.S. et al. 2013, MNRAS, 432, 121
\ref Pustilnik, S.; Kniazev, A.; Pramskij, A.; Izotov, Y.; Foltz, C.; Brosch, N.; Martin, J.-M.; Ugryumov, A.	
 2004, A\&A, 419, 469	
\ref Ramos Almeida, C., Rodr\'{i}guez Espinosa, J.M., Acosta-Pulido, J.A. Alonso-Herrero, A., 
P\'{e}rez Garc\'{i}a, A.M, Rodr\'{i}guez-Eugenio, N. 2013, MNRAS, 429, 3449
\ref Santini, P. et al. 2012, A\&A, 540, 39
\ref Schirmer, M., Diaz, R., Holhjem, K., Levenson, N.A., Winge, C. 2013, ApJ, 763, 60
\ref Seaton, M.J. 1975, MNRAS, 170, 475
\ref Shim, H., Colbert, J., Teplitz, H., Henry, A., Malkan, M., McCarthy, P., Yan, L. 2009, ApJ, 696, t85
\ref Spaans, M. \& Carollo, C.M. 1997 ApJ, 482, L93 
\ref Springel, V., Di Matteo T., Hernsquist, L. 2005, ApJ, 620, L79
\ref Swinbank, A.M., Smail, I., Sobral, D., Theuns, T., Best, P.N., Geach, J.E. ApJ, 2012, 760, 130)
\ref Stierwalt, S. et al 2013, ApJS, 206, 1
\ref Straughn, A.N. et al. 2009, 138, 1022
\ref Teplitz, H.I., Collins, N.R.,  Gardner, J.P., Hill, R.S., Rhodes, J 2003, ApJ, 589, 704
\ref Torres-Peimbert,S., Peimbert, M., Fierro, J. 1989, ApJ, 345, 186
\ref Veilleux, S., Kim, D.-C., Sanders, D.B. 1999 ApJ, 522, 113
\ref Viegas, S.M., Contini, M., Contini, T. 1999, A\&A, 347, 112
\ref Vila Costas, M.B., Edmunds, M.G. 1993, MNRAS, 265, 199
\ref Winter, L.M., Lewis, K.T., Koss, M., Veilleux, S., Keeney, B., Muschotzky, R. 
2010, ApJ, 710, 503
\ref Xia, L. et al.  2012, AJ, 144, 28

\end{document}